\documentclass[twoside,11pt]{article}

\usepackage{jmlr2e}
\usepackage{amsmath}
\usepackage{algorithm}
\usepackage{algorithmic}
\usepackage{subfigure}
\usepackage{graphicx}
\usepackage{bbm}
\usepackage{tikz}
\usetikzlibrary{arrows}
\usetikzlibrary{graphs}
\usepackage{url}

\ShortHeadings{IFM LAB TUTORIAL SERIES \# 4, COPYRIGHT \copyright IFM LAB}{JIAWEI ZHANG, IFM LAB DIRECTOR}
\firstpageno{1}

\begin{document}

\title{Secrets of the Brain: An Introduction to the Brain Anatomical Structure and Biological Function}

\author{\name Jiawei Zhang \email jiawei@ifmlab.org \\
	\addr{Founder and Director}\\
       {Information Fusion and Mining Laboratory}\\
       (First Version: April 2019; Revision: April 2019.)}

\maketitle

\begin{abstract}

In this paper, we will provide an introduction to the brain structure and function. Brain is an astonishing living organ inside our heads, weighing about 1.5$kg$, consisting of billions of tiny cells. The brain enables us to sense the world around us (to touch, to smell, to see and to hear, etc.), to think and to respond to the world as well. The main obstacles that prevent us from creating a machine which can behavior like real-world creatures are due to our limited knowledge about the brain in both its structure and its function. In this paper, we will focus introducing the brain anatomical structure and biological function, as well as its surrounding sensory systems. Many of the materials used in this paper are from wikipedia and several other neuroscience introductory articles, which will be properly cited in this article. This is the first of the three tutorial articles about the brain (the other two are \cite{zhang2019basic} and \cite{zhang2019cognitive}). In the follow-up two articles, we will further introduce the low-level composition basis structures (e.g., neuron, synapse and action potential) and the high-level cognitive functions (e.g., consciousness, attention, learning and memory) of the brain, respectively.

\end{abstract}


\tableofcontents

\section{Introduction}

\subsection{Nervous System}

The brain is a complex nervous system owned by animals, which can steer animals' movement and reactions in the natural environment. Generally, brain receives information from the external natural world via the \textit{sensory nerves}, which will be integrated and processed at the brain, and the reaction information will be sent to the body via the motor nerves. Almost all the animals on the planet have a brain or a similar nervous system structure that can function like a brain. We will provide an example as follows to illustrate the nervous systems owned by different types of animals on the earth, which structure complexity can vary from each other a lot.

\begin{example}
\begin{figure}[t]
    \centering
    \includegraphics[width=1.0\textwidth]{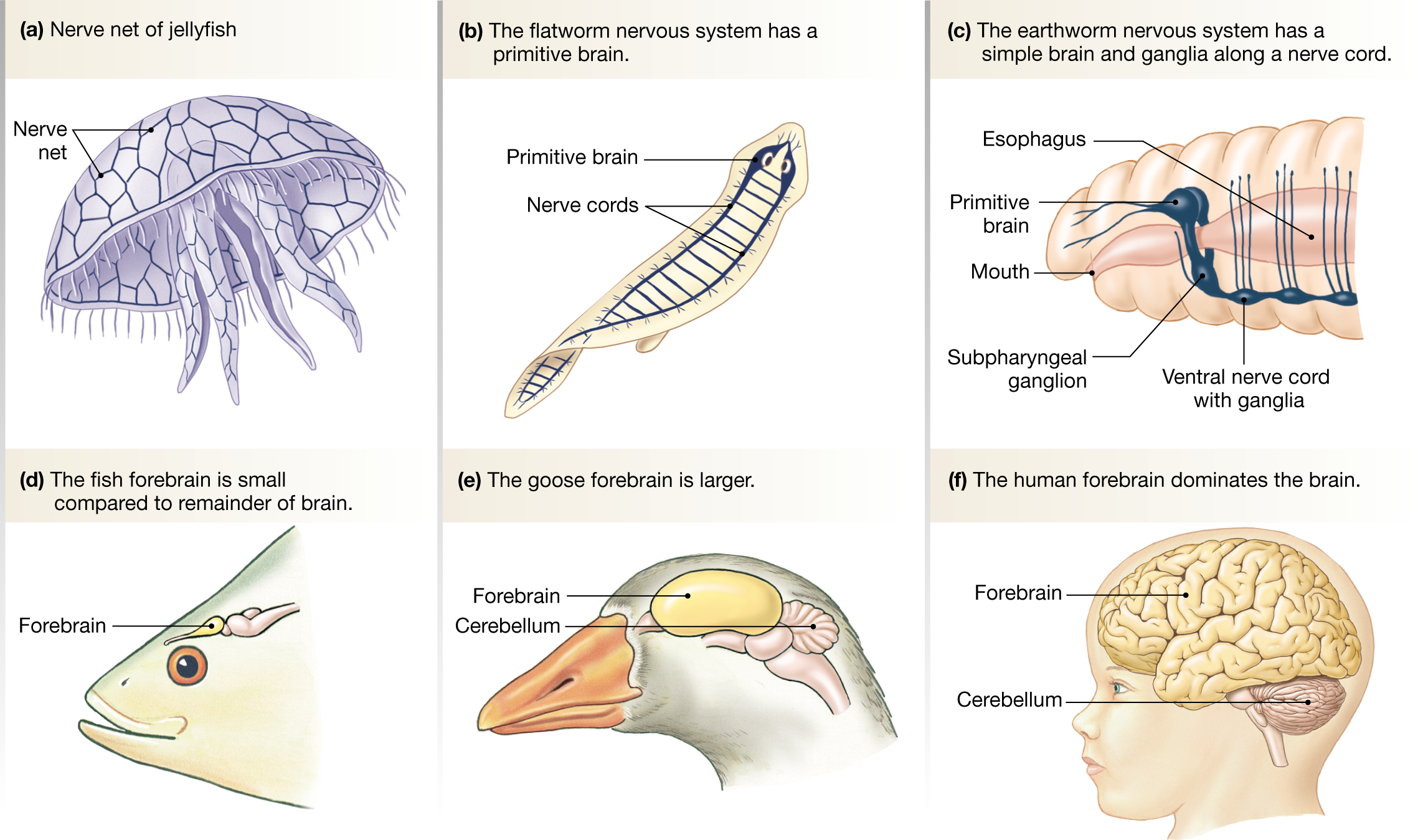}
    \caption{Examples of Animal Nervous Systems \cite{nervous_system}.}
    \label{fig:animal_brain_system}
\end{figure}

For instance, based on the introduction provided in \cite{nervous_system}, in Figure~\ref{fig:animal_brain_system}, we illustrate the nervous systems of several different creatures on the earth, including the \textit{jellyfish}, \textit{flatworm}, \textit{earthworm}, \textit{fish}, \textit{goose} and \textit{human beings}. 

\begin{itemize}

\item Some of the first multicellular animals to develop the neurons were members of the phylum Cnidaria, e.g., the jellyfish and the sea anemones. These animals have a nerve net consisting of sensory neurons, connective interneurons and motor neurons that innervate muscles and glands. These animals respond to stimuli with complex behaviors, yet without input from an identifiable control center.

\item In the primitive flatworms, we see the beginnings of a nervous system as we know it in higher animals, although in flatworms the distinction between central nervous system and peripheral nervous system is not clear. Flatworms have a rudimentary brain consisting of a cluster of nerve cell bodies concentrated in the head, or cephalic region. Two large nerves called nerve cords come off the primitive brain and lead to a nerve network that innervates distal regions of the flatworm body.

\item The segmented worms, or annelids, such as the earthworm, have a more advanced central nervous system. Clusters of cell bodies are no longer restricted to the head region, as they are in flatworms, but also occur in fused pairs, called ganglia, along a nerve cord. Because each segment of the worm contains a ganglion, simple reflexes can be integrated within a segment without input from the brain. Reflexes that do not require integration in the brain also occur in higher animals and are called spinal reflexes in humans and other vertebrates.

\item In vertebrate brain evolution, the most dramatic change is seen in the forebrain region, which includes the cerebrum. In fish, the forebrain is a small bulge dedicated mainly to processing olfactory information about odors in the environment. In birds and rodents, part of the forebrain has enlarged into a cerebrum with a smooth surface. In humans, the cerebrum is the largest and most distinctive part of the brain, with deep grooves and folds. More than anything else, the cerebrum is what makes us human. All evidence indicates that it is the part of the brain that allows reasoning and cognition.

\end{itemize}
\end{example}

\begin{figure}[t]
    \centering
    \includegraphics[width=0.6\textwidth]{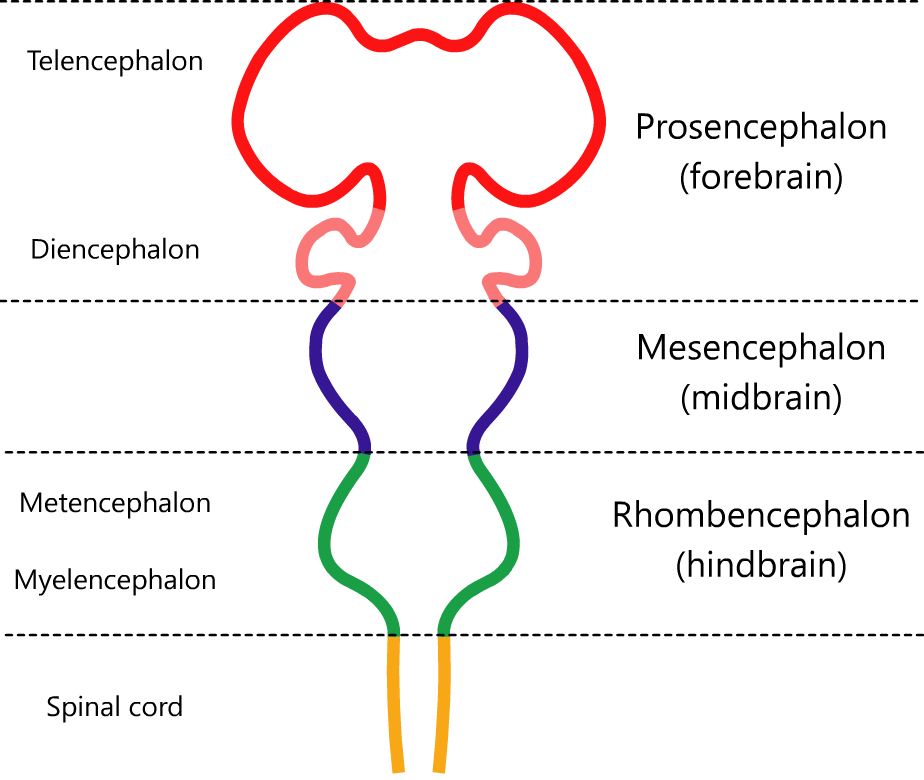}
    \caption{An Examples of Vertebrate Brain Structure \cite{brain_structure}.}
    \label{fig:animal_brain_structure}
\end{figure}

Even though the nervous systems structure differs greatly among different animals, they all function in a similar way based on signals passing through the neurons. Electrical signals in the form of action potentials, and chemical signals passing across synapses, are the same in all animals. It is only in the number and organization of the neurons that one species differs from another. Detailed information about the \textit{neuron}, \textit{synapse} and \textit{action potential} concepts are not covered in this paper, which will be introduced in detail in the upcoming tutorial article instead.

\subsection{Vertebrate Brain Structure}

All vertebrate brains share a common underlying form, which appears most clearly during early stages of embryonic development. In its earliest form, as shown in Figure~\ref{fig:animal_brain_structure}, the brain appears as three swellings at the front end of the neural tube; these swellings eventually become the \textit{forebrain}, \textit{midbrain}, and \textit{hindbrain} (or the \textit{prosencephalon}, \textit{mesencephalon}, and \textit{rhombencephalon}, respectively). At the earliest stages of brain development, the three areas are roughly equal in size. In many classes of vertebrates, such as fish and amphibians, the three parts remain similar in size in the adult, but in mammals the forebrain becomes much larger than the other parts, and the midbrain becomes very small.

For the mature vertebrates, the brain structure will be slightly different. The neuroanatomists usually divide the mature vertebrate brain into six main regions: the \textit{telencephalon} (\textit{cerebral hemispheres}), \textit{diencephalon} (\textit{thalamus} and \textit{hypothalamus}), \textit{mesencephalon} (\textit{midbrain}), \textit{cerebellum}, \textit{pons}, and \textit{medulla oblongata}. Each of these areas has a complex internal structure. Some parts, such as the \textit{cerebral cortex} and the \textit{cerebellar cortex}, consist of layers that are folded or convoluted to fit within the available space. Other parts, such as the \textit{thalamus} and \textit{hypothalamus}, consist of clusters of many small nuclei. Thousands of distinguishable areas can be identified within the vertebrate brain based on fine distinctions of neural structure, chemistry, and connectivity. This is a rough division of the vertebrate brain, and a more fine-granularity division of the human brain will be introduced in the following subsection.

\begin{figure}[t]
    \centering
    \includegraphics[width=0.6\textwidth]{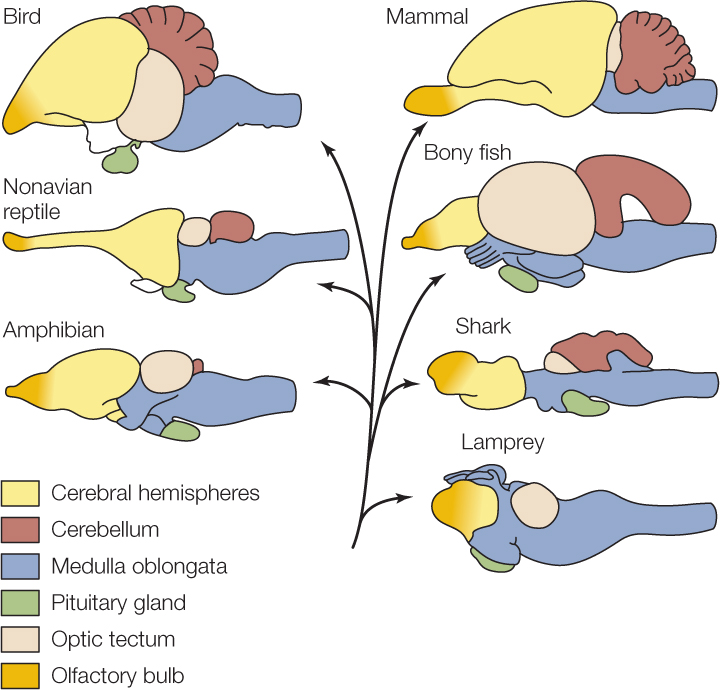}
    \caption{An Evolution of Vertebrate Animal Brains \cite{brain_comparison}.}
    \label{fig:animal_brain_comparison}
\end{figure}

\begin{example}

For instance, in Figure~\ref{fig:animal_brain_comparison}, we provide an evolution plot of vertebrate brain. A reasonably representative brain of a modern species in each group is shown. According to the plot, the composition of different types of vertebrate brains are quite similar to each other. In contrast, the cerebral hemispheres, which are a major component of the forebrain, have undergone dramatic changes during vertebrate evolution. 

There are two take-home messages from these comparisons. First, mammals and birds have evolved strikingly large cerebral hemispheres, the neurons and glia of which are of enormous importance in high-order brain function. Second, brain size matters. The evolution of enhanced functionality in mammals and birds has gone hand in hand with large increases in the numbers of neurons. From this vantage, it is significant that primates and dolphins have the largest brains, for their body sizes, of all mammals.

\end{example}

Although the same basic components are present in all vertebrate brains, some branches of vertebrate evolution have led to substantial distortions of brain geometry, especially in the forebrain area. The brain of a shark shows the basic components in a straightforward way, but in teleost fishes (the great majority of existing fish species), the forebrain has become ``everted'', like a sock turned inside out. In birds, there are also major changes in forebrain structure. These distortions can make it difficult to match brain components from one species with those of another species.

\section{Human Brain Anatomy}

The human brain can be divided into hundreds of parts, and in this part we will introduce several anatomical regions (including their structures, locations and functions) following the standard neuroanatomy hierarchies, including \textit{brainstem}, \textit{cerebellum}, \textit{diencephalon} and \textit{cerebrum}.

\begin{figure}[t]
    \centering
    \includegraphics[width=0.7\textwidth]{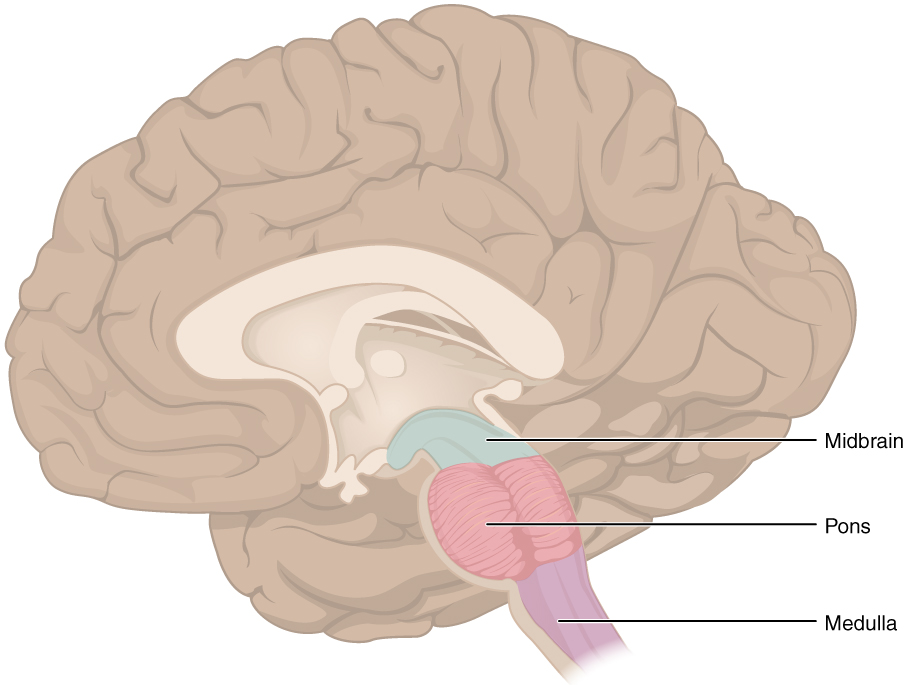}
    \caption{An Illustration of the Brainstem \cite{brain_stem}.}
    \label{fig:human_brain_stem}
\end{figure}

\subsection{Brainstem}

As illustrated in Figure~\ref{fig:human_brain_stem}, the brainstem (or brain stem) \cite{brain_stem} is the posterior part of the brain, continuous with the spinal cord. In the human brain the brainstem includes the midbrain, and the pons and medulla oblongata of the hindbrain. Sometimes the diencephalon, the caudal part of the forebrain, is included. The brainstem provides the main motor and sensory nerve supply to the face and neck via the cranial nerves. Of the thirteen pairs of cranial nerves, ten pairs (or twelve, if the diencephalon is included in the brainstem) come from the brainstem. The brainstem also plays an important role in the regulation of cardiac and respiratory function. It also regulates the central nervous system, and is pivotal in maintaining consciousness and regulating the sleep cycle. The brainstem has many basic functions including heart rate, breathing, sleeping, and eating.

\begin{itemize}
	\item \textbf{Pons}: The pons is part of the brainstem, and in humans and other bipeds lies inferior to the midbrain, superior to the medulla oblongata and anterior to the cerebellum. The pons in humans measures about 2.5 cm (0.98 in) in length. Most of it appears as a broad anterior bulge above the medulla. The pons contains nuclei that relay signals from the forebrain to the cerebellum, along with nuclei that deal primarily with sleep, respiration, swallowing, bladder control, hearing, equilibrium, taste, eye movement, facial expressions, facial sensation, and posture. The pons is implicated in sleep paralysis, and may also play a role in generating dreams.
	
	\item \textbf{Medulla Oblongata}: The medulla oblongata (or medulla) is a long stem-like structure located in the brainstem. It is anterior and partially inferior to the cerebellum. It is a cone-shaped neuronal mass responsible for autonomic (involuntary) functions ranging from vomiting to sneezing. The medulla contains the cardiac, respiratory, vomiting and vasomotor centers and therefore deals with the autonomic functions of breathing, heart rate and blood pressure.
	
	\item \textbf{Midbrain (Mesencephalon)}: The principal regions of the midbrain are the tectum, the cerebral aqueduct, tegmentum, and the cerebral peduncles. Rostrally the midbrain adjoins the diencephalon (thalamus, hypothalamus, etc.), while caudally it adjoins the hindbrain (pons and cerebellum). The midbrain is identified to be associated with vision, hearing, motor control, sleep/wake, arousal (alertness), and temperature regulation.
\end{itemize}

\begin{figure}[t]
    \centering
    \includegraphics[width=0.8\textwidth]{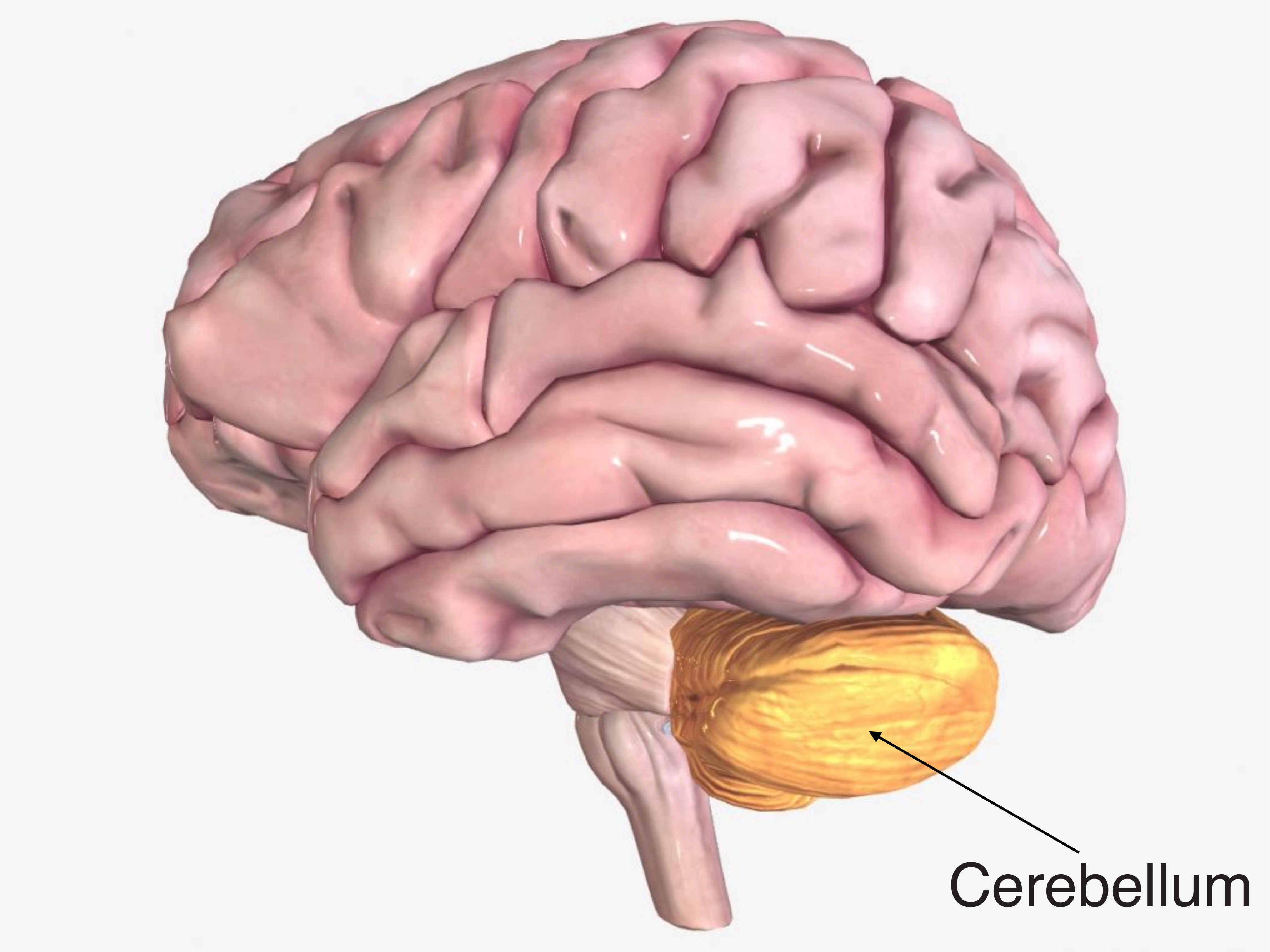}
    \caption{An Illustration of the Cerebellum \cite{brain_cerebellum}.}
    \label{fig:human_brain_cerebellum}
\end{figure}

\subsection{Cerebellum}

The cerebellum is a major feature of the hindbrain of all vertebrates. Although usually smaller than the cerebrum, in some animals such as the mormyrid fishes it may be as large as or even larger than the cerebrum.  In humans, the cerebellum plays an important role in motor control. It may also be involved in some cognitive functions such as attention and language as well as in regulating fear and pleasure responses, but its movement-related functions are the most solidly established. The human cerebellum does not initiate movement, but contributes to coordination, precision, and accurate timing: it receives input from sensory systems of the spinal cord and from other parts of the brain, and integrates these inputs to fine-tune motor activity. Cerebellar damage produces disorders in fine movement, equilibrium, posture, and motor learning in humans. In addition to its direct role in motor control, the cerebellum is necessary for several types of motor learning, most notably learning to adjust to changes in sensorimotor relationships. Several theoretical models have been developed to explain sensorimotor calibration in terms of synaptic plasticity within the cerebellum.\\

\begin{figure}[t]
    \vspace{-30pt}
    \centering
    \includegraphics[width=0.8\textwidth]{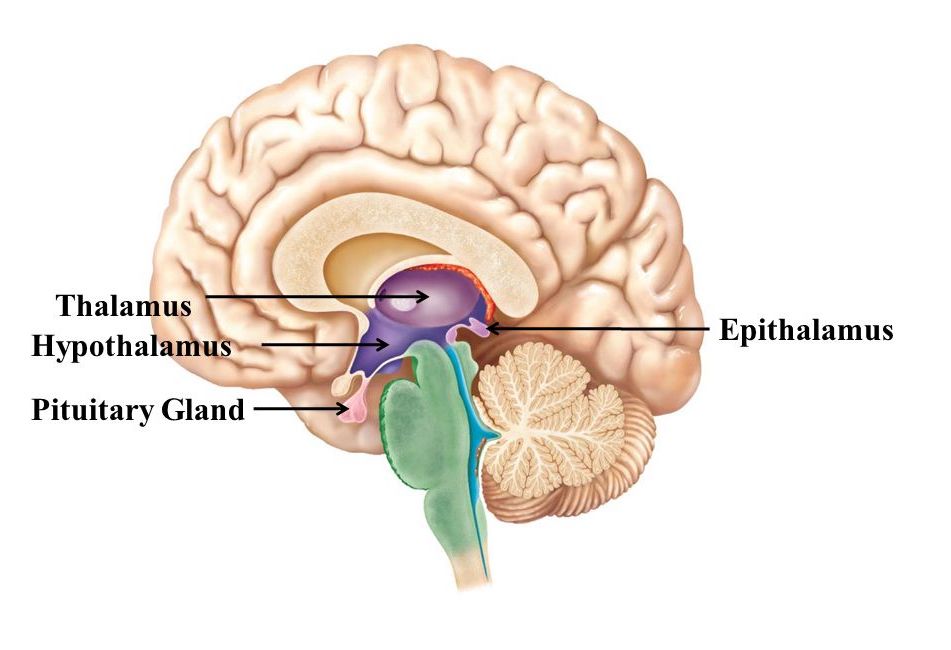}
    \vspace{-30pt}
    \caption{An Illustration of the Diencephalon \cite{brain_diencephalon}.}
    \label{fig:human_brain_diencephalon}
\end{figure}

\subsection{Diencephalon}

The diencephalon is a division of the forebrain, and is situated between the telencephalon and the midbrain \cite{brain_diencephalon_description}. As shown in Figure~\ref{fig:human_brain_diencephalon}, it consists of structures that are on either side of the third ventricle, including the thalamus, the hypothalamus (including the posterior pituitary), the epithalamus and the subthalamus. Despite being a relatively small part of the central nervous system in terms of mass, the diencephalon plays a number of critical roles in healthy brain and bodily function, from regulating wakefulness to controlling the autonomic nervous system.

\begin{itemize}
	\item \textbf{Thalamus}: The thalamus consists of two oval collections of nuclei that make up most of the mass of the diencephalon. The thalamus is often described as a relay station because almost all sensory information (with the exception of smell) that proceeds to the cortex first stops in the thalamus before being sent on to its destination. The structure is subdivided into a number of nuclei that possess functional specializations for dealing with particular types of information. Sensory information thus travels to the thalamus and is routed to a nucleus tailored to dealing with that type of sensory data. Then, the information is sent from that nucleus to the appropriate area in the cortex where it is further processed.
	
	The thalamus doesn't deal just with sensory information, however. It also receives a great deal of information from the cerebral cortex, and it is involved with processing that information and sending it back to other areas of the brain. Due to its involvement in these complex networks, the thalamus plays a role in a number of important functions ranging from sleep to consciousness.

	\item \textbf{Hypothalamus}: The hypothalamus is a small (about the size of an almond) region located directly above the brainstem. It also is made up of a collection of nuclei that are involved in a variety of functions. The hypothalamus is often linked, however, to two main roles: maintaining homeostasis and regulating hormone release. 
	
	Homeostasis is the maintenance of equilibrium in a system like the human body. Optimal biological function is facilitated by keeping things like body temperature, blood pressure, and caloric intake/expenditure at a fairly constant level. The hypothalamus receives a steady stream of information about these types of factors. When it recognizes an unanticipated imbalance, it enacts a mechanism to rectify that disparity.
	
	The hypothalamus acts to maintain homeostasis---and influences behavior in general---by regulating hormone secretion. This is primarily done through the control of hormone release from the pituitary gland. Through this mechanism, the hypothalamus has widespread effects on the body and behavior. It is often said that the hypothalamus is responsible for the four Fs: fighting, fleeing, feeding, and fornication. Clearly, due to the frequency and significance of these behaviors, the hypothalamus is extremely important in everyday life.
	
\begin{figure}[t]
    \vspace{-30pt}
    \centering
    \includegraphics[width=0.8\textwidth]{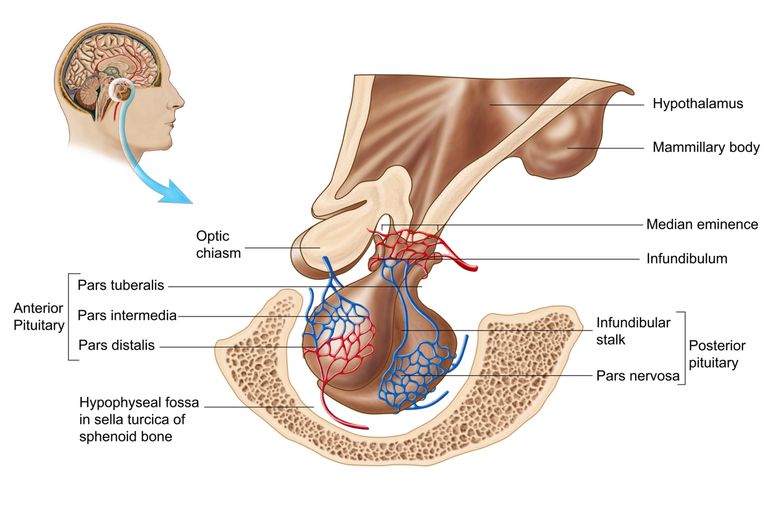}
    \vspace{-20pt}
    \caption{An Illustration of the Pituitary Gland \cite{pituitary_gland_plot}.}
    \label{fig:human_brain_pituitary_gland}
\end{figure}

	\item \textbf{Epithalamus}: The epithalamus consists primarily of the pineal gland and the habenulae. The pineal gland is an endocrine gland that secretes the hormone melatonin, which is thought to play an important role in the regulation of circadian rhythms. The habenulae (more often referred to with the singular: habenula) are two small areas near the pineal gland. The functions of the habenula are poorly understood, but it is thought to potentially be involved with reward processing and has been implicated in depression. Additionally, there is some evidence that the habenula also produces melatonin, and that it might be involved with sleep regulation.
	
	\item \textbf{Subthalamus}: A portion of the subthalamus is made up of tissue from the midbrain extending into the diencephalon. Thus, parts of midbrain regions like the substantia nigra and red nucleus are found in the diencephalon. The subthalamus is also home to the subthalamic nucleus and the zona incerta. The subthalamic nucleus is densely interconnected with the basal ganglia, and plays a role in modulating movement. The zona incerta has many connections throughout the cortex and spinal cord, but its function is still not determined. Several collections of important fibers (e.g. somatosensory fibers) also pass through the subthalamus.

	\item \textbf{Pituitary Gland}: The pituitary gland \cite{pituitary_gland}, in humans, is a pea-sized gland that sits in a protective bony enclosure called the sella turcica. As illustrated in Figure~\ref{fig:human_brain_pituitary_gland}, it is composed of three lobes: anterior, intermediate, and posterior. In many animals, these three lobes are distinct.The intermediate is avascular and almost absent in human beings. The intermediate lobe is present in many lower animal species, in particular in rodents, mice and rats, that have been used extensively to study pituitary development and function. In all animals, the fleshy, glandular anterior pituitary is distinct from the neural composition of the posterior pituitary, which is an extension of the hypothalamus.
	\begin{itemize}
		\item \textbf{Anterior Pituitary}: The anterior pituitary arises from an invagination of the oral ectoderm and forms Rathke's pouch. This contrasts with the posterior pituitary, which originates from neuroectoderm. Endocrine cells of the anterior pituitary are controlled by regulatory hormones released by parvocellular neurosecretory cells in the hypothalamic capillaries leading to infundibular blood vessels, which in turn lead to a second capillary bed in the anterior pituitary. The anterior pituitary synthesizes and secretes hormones, including growth hormone, corticotropin-releasing hormone, Thyroid-stimulating hormone, Luteinizing hormone and Follicle-stimulating hormone, etc.
		\item \textbf{Posterior Pituitary}: The posterior lobe develops as an extension of the hypothalamus. The posterior pituitary hormones are synthesized by cell bodies in the hypothalamus. The magnocellular neurosecretory cells, of the supraoptic and paraventricular nuclei located in the hypothalamus, project axons down the infundibulum to terminals in the posterior pituitary. This simple arrangement differs sharply from that of the adjacent anterior pituitary, which does not develop from the hypothalamus. The posterior pituitary stores and secretes (but does not synthesize) the following important endocrine hormones: Antidiuretic hormone and Oxytocin, etc.
	\end{itemize}
\end{itemize}

\begin{figure}[t]
    \centering
    \includegraphics[width=1.0\textwidth]{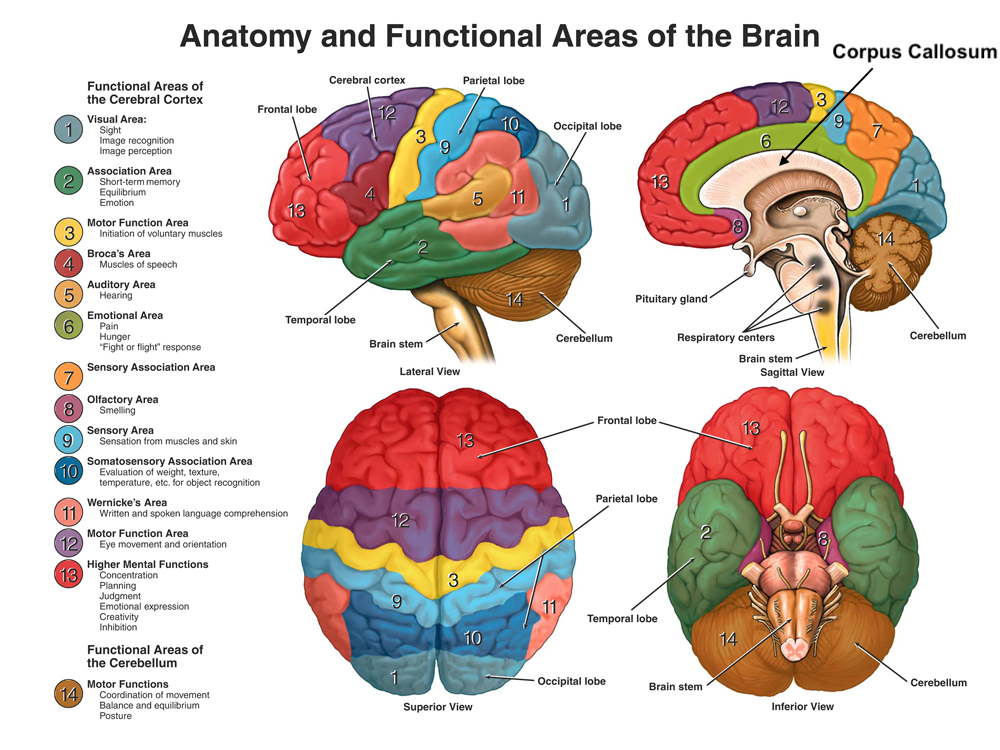}
    \caption{An Illustration of Human Brain Cerebrum \cite{brain_amatomy}.}
    \label{fig:human_brain_cerebrum}
\end{figure}

\subsection{Cerebrum}

The cerebrum (as illustrated in Figure~\ref{fig:human_brain_cerebrum}) is a largest part of the human brain containing the cerebral cortex (of the two cerebral hemispheres), as well as several subcortical structures, including the hippocampus, basal ganglia, and olfactory bulb. In the human brain, the cerebrum is the uppermost region of the central nervous system. The prosencephalon is the embryonic structure from which the cerebrum develops prenatally. In mammals, the dorsal telencephalon, or pallium, develops into the cerebral cortex, and the ventral telencephalon, or subpallium, becomes the basal ganglia. The cerebrum is also divided into approximately symmetric left and right cerebral hemispheres. It functions as the center of sensory perception, memory, thoughts and judgement; the cerebrum also functions as the center of voluntary motor activities.

\begin{itemize}

	\item \textbf{Corpus Callosum}: The human cerebrum can be divided into the left and right cerebral hemispheres, which are connected by the corpus callosum, as highlighted in the top right plot of Figure~\ref{fig:human_brain_cerebrum}. The corpus callosum, also callosal commissure, is a wide, thick, nerve tract consisting of a flat bundle of commissural fibers, beneath the cerebral cortex in the brain. The corpus callosum is only found in placental mammals. It spans part of the longitudinal fissure, connecting the left and right cerebral hemispheres, enabling communication between them. It is the largest white matter structure in the human brain, about ten centimetres in length and consisting of 200-300 million axonal projections.
	
	\item \textbf{Cerebral Cortex}: The cerebral cortex (plural cortices), also known as the cerebral mantle, is the outer layer of neural tissue of the cerebrum of the brain, in humans and other mammals. It is separated into two cortices, by the longitudinal fissure that divides the cerebrum into the left and right cerebral hemispheres. The two hemispheres are joined beneath the cortex by the corpus callosum. The cerebral cortex is the largest site of neural integration in the central nervous system. It plays a key role in memory, attention, perception, awareness, thought, language, and consciousness.

In most mammals, apart from small mammals that have small brains, the cerebral cortex is folded, providing a greater surface area in the confined volume of the cranium. Apart from minimizing brain and cranial volume cortical folding is crucial for the wiring of the brain and its functional organization. In mammals with a small brain there is no folding and the cortex is smooth. There are between 14 and 16 billion neurons in the cerebral cortex. These are organized into cortical columns and minicolumns of neurons that make up the layers of the cortex. Most of the cerebral cortex consists of the six-layered neocortex. The cerebral cortex can be divided into four main lobes, as indicated by the four different colors in Figure~\ref{fig:human_brain_lobe}. The lobes of the brain were originally a purely anatomical classification, but have been shown also to be related to different brain functions.

\begin{figure}[t]
    \centering
    \includegraphics[width=0.8\textwidth]{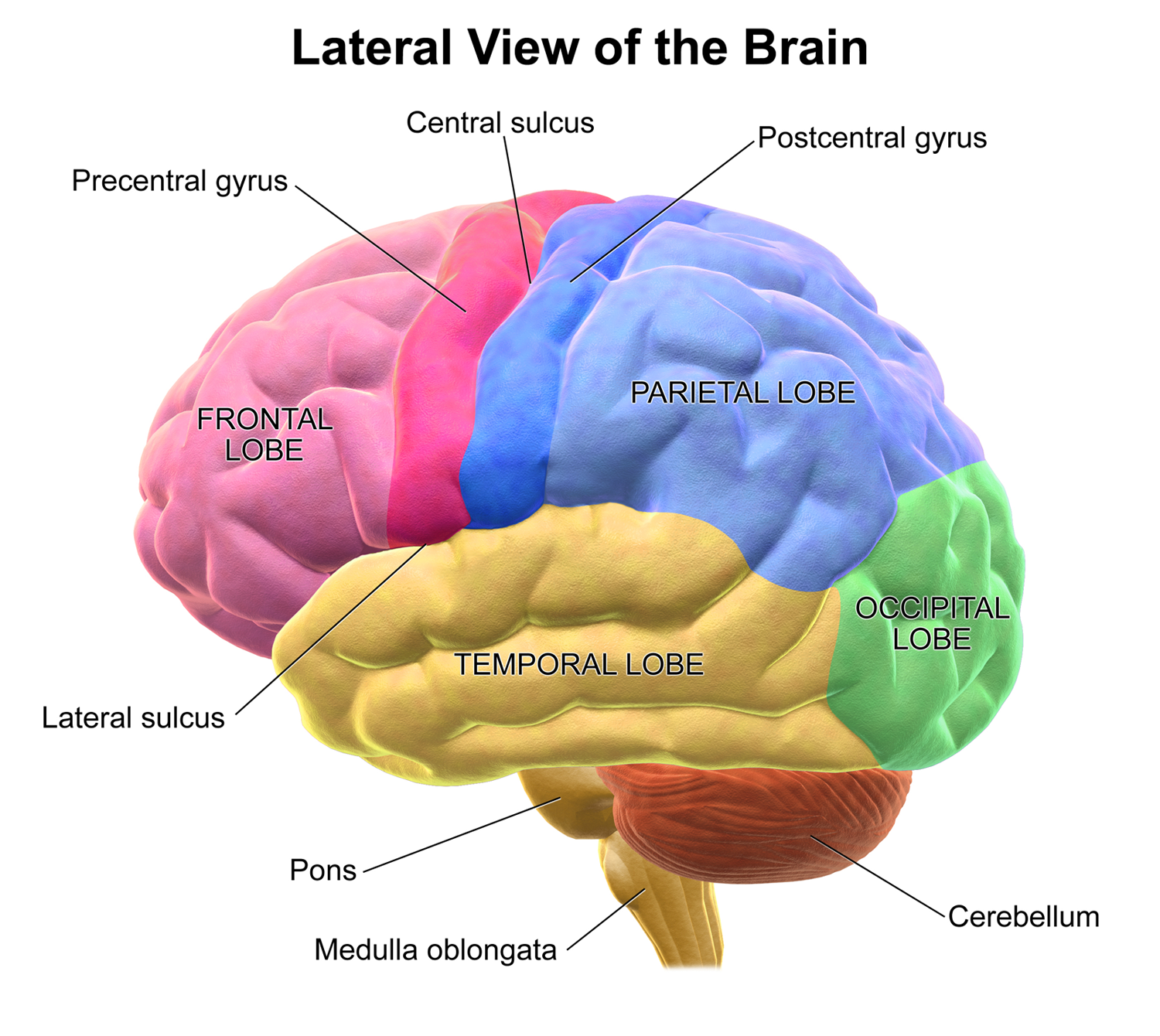}
    \caption{An Illustration of Human Brain Lobes.}
    \label{fig:human_brain_lobe}
\end{figure}

	\begin{itemize}
	
		\item \textbf{Frontal Lobe}: The frontal lobe is the largest of the four major lobes of the brain in mammals, and is located at the front of each hemisphere (in front of the parietal lobe and the temporal lobe). It is separated from the parietal lobe by a groove between tissues called the central sulcus, and from the temporal lobe by a deeper groove called the lateral sulcus. The frontal lobe is covered by the frontal cortex. The frontal cortex includes the premotor cortex, and the primary motor cortex, i.e., the cortical parts of the motor cortex.
		
		\begin{itemize}
		
			\item \textbf{Motor Cortex}: The motor cortex is the region of the cerebral cortex involved in the planning, control, and execution of voluntary movements. Classically the motor cortex is an area of the frontal lobe located in the posterior precentral gyrus immediately anterior to the central sulcus. The motor cortex can be divided into three areas:
			\begin{enumerate}
				\item The primary motor cortex is the main contributor to generating neural impulses that pass down to the spinal cord and control the execution of movement. However, some of the other motor areas in the brain also play a role in this function. It is located on the anterior paracentral lobule on the medial surface.
				\item The premotor cortex is responsible for some aspects of motor control, possibly including the preparation for movement, the sensory guidance of movement, the spatial guidance of reaching, or the direct control of some movements with an emphasis on control of proximal and trunk muscles of the body. Located anterior to the primary motor cortex.
				\item The supplementary motor area (or SMA), has many proposed functions including the internally generated planning of movement, the planning of sequences of movement, and the coordination of the two sides of the body such as in bi-manual coordination. Located on the midline surface of the hemisphere anterior to the primary motor cortex.
			\end{enumerate}
			
\begin{figure}[t]
    \centering
    \includegraphics[width=0.7\textwidth]{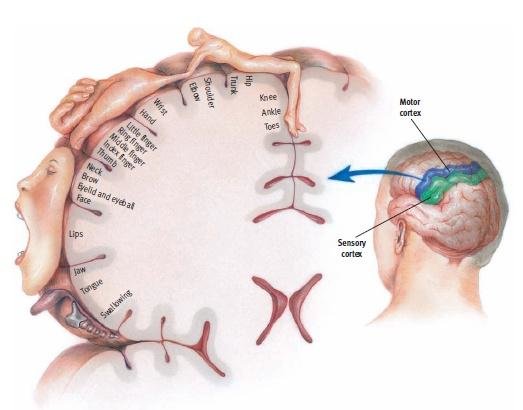}
    \caption{An Illustration of motor homunculus in primary motor cortex \cite{phdthesis}.}
    \label{fig:human_brain_motor_control}
\end{figure}

			The brain hemispheres control the body in an interesting way. Signals from the motor cortex cross the body's midline to activate skeletal muscles on the opposite side of the body, meaning that the left hemisphere of the brain controls the right side of the body, and the right hemisphere controls the left side of the body. Every part of the body is represented in the primary motor cortex, and these representations are arranged somatotopically, i.e., the foot is next to the leg which is next to the trunk which is next to the arm and the hand. The amount of brain matter devoted to any particular body part represents the amount of control that the primary motor cortex has over that body part. For example, as indicated in Figure~\ref{fig:human_brain_motor_control}, a lot of cortical space is required to control the complex movements of the hand and fingers, and these body parts have larger representations in motor cortex than the trunk or legs, whose muscle patterns are relatively simple. This disproportionate map of the body in the motor cortex is called the motor homunculus.
			
		\end{itemize}

		\item \textbf{Parietal Lobe}: The parietal lobe is positioned above the temporal lobe and behind the frontal lobe and central sulcus. The parietal lobe integrates sensory information among various modalities, including spatial sense and navigation (proprioception), the main sensory receptive area for the sense of touch (mechanoreception) in the somatosensory cortex which is just posterior to the central sulcus in the postcentral gyrus, and the dorsal stream of the visual system. The major sensory inputs from the skin (touch, temperature, and pain receptors), relay through the thalamus to the parietal lobe.
		
		\begin{itemize}
		
			\item \textbf{Sensory Cortex}: The sensory cortex can refer informally to the primary somatosensory cortex. The primary somatosensory cortex is located in the postcentral gyrus, and is part of the somatosensory system. At the primary somatosensory cortex, tactile representation is orderly arranged (in an inverted fashion) from the toe (at the top of the cerebral hemisphere) to mouth (at the bottom). However, some body parts may be controlled by partially overlapping regions of cortex. Each cerebral hemisphere of the primary somatosensory cortex only contains a tactile representation of the opposite (contralateral) side of the body. The amount of primary somatosensory cortex devoted to a body part is not proportional to the absolute size of the body surface, but, instead, to the relative density of cutaneous tactile receptors on that body part. The density of cutaneous tactile receptors on a body part is generally indicative of the degree of sensitivity of tactile stimulation experienced at said body part. For this reason, the human lips and hands have a larger representation than other body parts.
			
		\end{itemize}
	
		\item \textbf{Occipital Lobe}: The occipital lobe is one of the four major lobes of the cerebral cortex in the brain of mammals. The occipital lobe is the visual processing center of the mammalian brain containing most of the anatomical region of the visual cortex. The primary visual cortex is Brodmann area 17 (the Brodmann areas will be introduced in Section~\ref{subsec:brodman_area}), commonly called V1 (i.e., visual one). Human V1 is located on the medial side of the occipital lobe within the calcarine sulcus; the full extent of V1 often continues onto the posterior pole of the occipital lobe. V1 is often also called striate cortex because it can be identified by a large stripe of myelin, the Stria of Gennari. Visually driven regions outside V1 are called extrastriate cortex. There are many extrastriate regions, and these are specialized for different visual tasks, such as visuospatial processing, color differentiation, and motion perception.
		
		\item \textbf{Temporal Lobe}: The temporal lobe lies between frontal lobe and occipital lobe, and below the parietal lobe. The temporal lobe is involved in processing sensory input into derived meanings for the appropriate retention of visual memory, language comprehension, and emotion association. The temporal lobe consists of structures that are vital for declarative or long-term memory. Declarative (denotative) or explicit memory is conscious memory divided into semantic memory (facts) and episodic memory (events). Medial temporal lobe structures that are critical for long-term memory include the hippocampus, along with the surrounding hippocampal region consisting of the perirhinal, parahippocampal, and entorhinal neocortical regions. The hippocampus is critical for memory formation, and the surrounding medial temporal cortex is currently theorized to be critical for memory storage. The prefrontal and visual cortices are also involved in explicit memory
		
	\end{itemize}
	
\end{itemize}

These brain cerebral cortex can be further divided into several regions, each of which is responsible for certain functions. As illustrated in Figure~\ref{fig:human_brain_cerebrum}, a rough division of the cerebral cortex functional areas is provided. In the following section, we will introduce a much more fine-granularity cerebral cortex functional area division, which is also referred to as the \textit{Brodmann area} formally.

\subsection{Brodmann Area}\label{subsec:brodman_area}

As introduced in \cite{brodmann_area}, different parts of the cerebral cortex are involved in different cognitive and behavioral functions. The differences show up in a number of ways: the effects of localized brain damage, regional activity patterns exposed when the brain is examined using functional imaging techniques, connectivity with subcortical areas, and regional differences in the cellular architecture of the cortex. Neuroscientists describe most of the cortex-the part they call the neocortex-as having six layers, but not all layers are apparent in all areas, and even when a layer is present, its thickness and cellular organization may vary. Scientists have constructed maps of cortical areas on the basis of variations in the appearance of the layers as seen with a microscope. One of the most widely used schemes came from Korbinian Brodmann, who split the cortex into 52 different areas and assigned each a number (many of these Brodmann areas have since been subdivided)

The Brodmann areas denote the regions of the cerebral cortex, in the human or other primate brain, defined by its cytoarchitecture, or histological structure and organization of cells. Brodmann areas were originally defined and numbered by the German anatomist Korbinian Brodmann based on the cytoarchitectural organization of neurons he observed in the cerebral cortex using the Nissl method of cell staining. Brodmann published his maps of cortical areas in humans, monkeys, and other species in 1909, along with many other findings and observations regarding the general cell types and laminar organization of the mammalian cortex. The same Brodmann area number in different species does not necessarily indicate homologous areas.

Brodmann areas have been discussed, debated, refined, and renamed exhaustively for nearly a century and remain the most widely known and frequently cited cytoarchitectural organization of the human cortex. Many of the areas Brodmann defined based solely on their neuronal organization have since been correlated closely to diverse cortical functions. For example, Brodmann areas 3, 1 and 2 are the primary somatosensory cortex; area 4 is the primary motor cortex; area 17 is the primary visual cortex; and areas 41 and 42 correspond closely to primary auditory cortex. Higher order functions of the association cortical areas are also consistently localized to the same Brodmann areas by neurophysiological, functional imaging, and other methods (e.g., the consistent localization of Broca's speech and language area to the left Brodmann areas 44 and 45). However, functional imaging can only identify the approximate localization of brain activations in terms of Brodmann areas since their actual boundaries in any individual brain requires its histological examination.

\begin{figure}[t]
    \centering
    \includegraphics[width=1.0\textwidth]{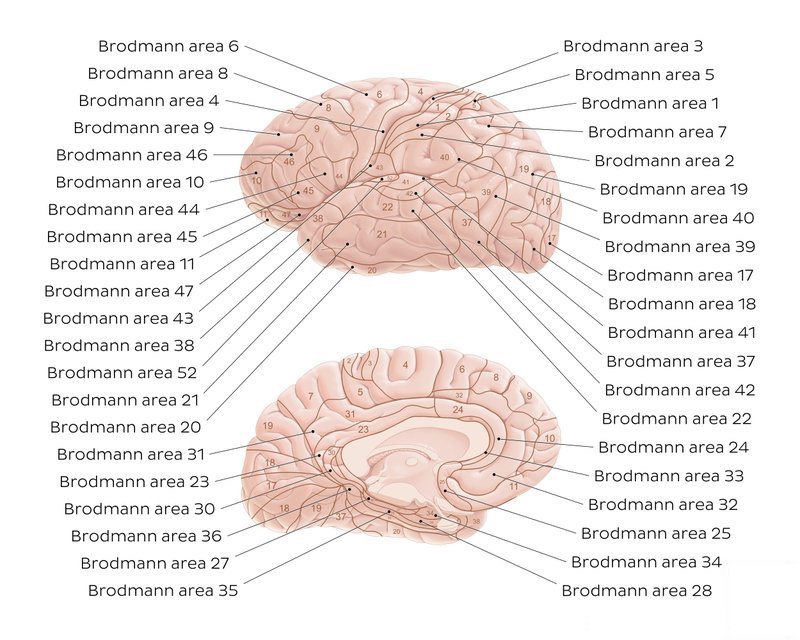}
    \caption{An Illustration of Human Cerebral Cortex Brodmann Areas (top: lateral surface; bottom: medical surface) \cite{brodmann_area_plot}.}
    \label{fig:human_brain_brodmann_area_plot}
\end{figure}

We also provide a list of the Brodmann areas of the human (and some other primates) brain cerebral cortex as follows. The locations of these areas visible on the lateral surface and the medial surface\footnote{The view of the section between the right and left hemispheres of the brain is denoted the ``medial surface''} are illustrated in Figure~\ref{fig:human_brain_brodmann_area_plot}.
\begin{itemize}
\item Areas 3, 1 and 2 - Primary somatosensory cortex in the postcentral gyrus (frequently referred to as Areas 3, 1, 2 by convention)
\item Area 4 - Primary motor cortex
\item Area 5 - Superior parietal lobule
\item Area 6 - Premotor cortex and Supplementary Motor Cortex (Secondary Motor Cortex) (Supplementary motor area)
\item Area 7 - Visuo-Motor Coordination
\item Area 8 - Includes Frontal eye fields
\item Area 9 - Dorsolateral prefrontal cortex
\item Area 10 - Anterior prefrontal cortex (most rostral part of superior and middle frontal gyri)
\item Area 11 - Orbitofrontal area (orbital and rectus gyri, plus part of the rostral part of the superior frontal gyrus)
\item Area 12 - Orbitofrontal area (used to be part of BA11, refers to the area between the superior frontal gyrus and the inferior rostral sulcus)
\item Area 13 and Area 14* - Insular cortex
\item Area 15* - Anterior Temporal lobe
\item Area 16 - Insular cortex
\item Area 17 - Primary visual cortex (V1)
\item Area 18 - Secondary visual cortex (V2)
\item Area 19 - Associative visual cortex (V3, V4, V5)
\item Area 20 - Inferior temporal gyrus
\item Area 21 - Middle temporal gyrus
\item Area 22 - Part of the superior temporal gyrus, included in Wernicke's area
\item Area 23 - Ventral posterior cingulate cortex
\item Area 24 - Ventral anterior cingulate cortex.
\item Area 25 - Subgenual area (part of the Ventromedial prefrontal cortex)
\item Area 26 - Ectosplenial portion of the retrosplenial region of the cerebral cortex
\item Area 27 - Piriform cortex
\item Area 28 - Ventral entorhinal cortex
\item Area 29 - Retrosplenial cortex
\item Area 30 - Subdivision of retrosplenial cortex
\item Area 31 - Dorsal Posterior cingulate cortex
\item Area 32 - Dorsal anterior cingulate cortex
\item Area 33 - Part of anterior cingulate cortex
\item Area 34 - Dorsal entorhinal cortex (on the Parahippocampal gyrus)
\item Area 35 - Part of the perirhinal cortex (in the rhinal sulcus)
\item Area 36 - Part of the perirhinal cortex (in the rhinal sulcus)
\item Area 37 - Fusiform gyrus
\item Area 38 - Temporopolar area (most rostral part of the superior and middle temporal gyri)
\item Area 39 - Angular gyrus, considered by some to be part of Wernicke's area
\item Area 40 - Supramarginal gyrus considered by some to be part of Wernicke's area
\item Areas 41 and 42 - Auditory cortex
\item Area 43 - Primary gustatory cortex
\item Areas 44 and 45 - Broca's area, includes the opercular part and triangular part of the inferior frontal gyrus
\item Area 46 - Dorsolateral prefrontal cortex
\item Area 47 - Orbital part of inferior frontal gyrus
\item Area 48 - Retrosubicular area (a small part of the medial surface of the temporal lobe)
\item Area 49 - Parasubicular area in a rodent
\item Area 52 - Parainsular area (at the junction of the temporal lobe and the insula)
\end{itemize}
As the readers may notice, the Area 50 and Area 51 are missing in the above list, which are missing in the human brain cerebral cortex and were not provided in Brodmann's original human map of the brain. Among all the above areas, some of the areas marked with $*$ (i.e., Area 14 and Area 15), and they are only found in non-human primates.

\section{Sensory System}\label{sec:sensory_system}

The sensory nervous system is a part of the nervous system responsible for processing sensory information. A sensory system consists of sensory neurons (including the sensory receptor cells), neural pathways, and parts of the brain involved in sensory perception. Commonly recognized sensory systems are those for vision, hearing, touch, taste, smell, and balance. In short, senses are transducers from the physical world to the realm of the mind where we interpret the information, creating our perception of the world around us.

\subsection{Stimulus and Receptor}

Sensory systems code for four aspects of a stimulus; type (modality), intensity, location, and duration. Arrival time of a sound pulse and phase differences of continuous sound are used for sound localization. Certain receptors are sensitive to certain types of stimuli (for example, different mechanoreceptors respond best to different kinds of touch stimuli, like sharp or blunt objects). Receptors send impulses in certain patterns to send information about the intensity of a stimulus (for example, how loud a sound is). The location of the receptor that is stimulated gives the brain information about the location of the stimulus (for example, stimulating a mechanoreceptor in a finger will send information to the brain about that finger). The duration of the stimulus (how long it lasts) is conveyed by firing patterns of receptors. These impulses are transmitted to the brain through afferent neurons.

While debate exists among neurologists as to the specific number of senses due to differing definitions of what constitutes a sense, Gautama Buddha and Aristotle classified five ``traditional'' human senses which have become universally accepted: touch, taste, smell, sight, and hearing. Other senses that have been well-accepted in most mammals, including humans, include nociception, equilibrioception, kinaesthesia, and thermoception. Furthermore, some nonhuman animals have been shown to possess alternate senses, including magnetoception and electroreception. 

The initialization of sensation stems from the response of a specific receptor to a physical stimulus. The receptors which react to the stimulus and initiate the process of sensation are commonly characterized in four distinct categories: chemoreceptors, photoreceptors, mechanoreceptors, and thermoreceptors. All receptors receive distinct physical stimuli and transduce the signal into an electrical action potential. This action potential then travels along afferent neurons to specific brain regions where it is processed and interpreted.

\begin{itemize}
\item \textbf{Chemoreceptors}: Chemoreceptors, or chemosensors, detect certain chemical stimuli and transduce that signal into an electrical action potential. The two primary types of chemoreceptors are:
\begin{itemize}
\item Distance chemoreceptors are integral to receiving stimuli in the olfactory system through both olfactory receptor neurons and neurons in the vomeronasal organ.
\item Direct chemoreceptors include the taste buds in the gustatory system as well as receptors in the aortic bodies which detect changes in oxygen concentration.
\end{itemize}

\item \textbf{Photoreceptors}: Photoreceptors are capable of phototransduction, a process which converts light (electromagnetic radiation) into, among other types of energy, a membrane potential. The three primary types of photoreceptors are: Cones are photoreceptors which respond significantly to color. In humans the three different types of cones correspond with a primary response to short wavelength (blue), medium wavelength (green), and long wavelength (yellow/red). Rods are photoreceptors which are very sensitive to the intensity of light, allowing for vision in dim lighting. The concentrations and ratio of rods to cones is strongly correlated with whether an animal is diurnal or nocturnal. In humans rods outnumber cones by approximately 20:1, while in nocturnal animals, such as the tawny owl, the ratio is closer to 1000:1. Ganglion Cells reside in the adrenal medulla and retina where they are involved in the sympathetic response. Of the ~1.3 million ganglion cells present in the retina, 1-2\% are believed to be photosensitive ganglia. These photosensitive ganglia play a role in conscious vision for some animals, and are believed to do the same in humans.

\item \textbf{Mechanoreceptors}: Mechanoreceptors are sensory receptors which respond to mechanical forces, such as pressure or distortion. While mechanoreceptors are present in hair cells and play an integral role in the vestibular and auditory systems, the majority of mechanoreceptors are cutaneous and are grouped into four categories:
\begin{itemize}
\item Slowly adapting type 1 receptors have small receptive fields and respond to static stimulation. These receptors are primarily used in the sensations of form and roughness.
\item Slowly adapting type 2 receptors have large receptive fields and respond to stretch. Similarly to type 1, they produce sustained responses to a continued stimuli.
\item Rapidly adapting receptors have small receptive fields and underlie the perception of slip.
\item Pacinian receptors have large receptive fields and are the predominant receptors for high-frequency vibration.
\end{itemize}

\item \textbf{Thermoreceptors}: Thermoreceptors are sensory receptors which respond to varying temperatures. While the mechanisms through which these receptors operate is unclear, recent discoveries have shown that mammals have at least two distinct types of thermoreceptors:
\begin{itemize}
\item The end-bulb of Krause, or bulboid corpuscle, detects temperatures above body temperature.
\item Ruffini's end organ detects temperatures below body temperature.
\end{itemize}

\item \textbf{Nociceptors}: Nociceptors respond to potentially damaging stimuli by sending signals to the spinal cord and brain. This process, called nociception, usually causes the perception of pain. They are found in internal organs, as well as on the surface of the body. Nociceptors detect different kinds of damaging stimuli or actual damage. Those that only respond when tissues are damaged are known as ``sleeping'' or ``silent'' nociceptors.
\begin{itemize}
\item Thermal nociceptors are activated by noxious heat or cold at various temperatures.
\item Mechanical nociceptors respond to excess pressure or mechanical deformation.
\item Chemical nociceptors respond to a wide variety of chemicals, some of which are signs of tissue damage. They are involved in the detection of some spices in food.
\end{itemize}

\end{itemize}

All stimuli received by the receptors listed above are transduced to an action potential, which is carried along one or more afferent neurons towards a specific area of the brain. These different receptors together with the corresponding sensory cortex will compose several different human sensory subsystems, including \textit{visual system}, \textit{auditory system}, \textit{somatosensory system}, \textit{taste system}, \textit{olfactory system} and \textit{vestibular system}, which will all be introduced in the follow part respectively.

\subsection{Visual System}

According to \cite{visual_system}, the visual system is the part of the central nervous system which gives organisms the ability to process visual detail as sight, as well as enabling the formation of several non-image photo response functions. It detects and interprets information from visible light to build a representation of the surrounding environment. The visual system carries out a number of complex tasks, including the reception of light and the formation of monocular representations; the buildup of a nuclear binocular perception from a pair of two dimensional projections; the identification and categorization of visual objects; assessing distances to and between objects; and guiding body movements in relation to the objects seen. The psychological process of visual information is known as visual perception, a lack of which is called blindness. Non-image forming visual functions, independent of visual perception, include the pupillary light reflex (PLR) and circadian photoentrainment.

\begin{figure}[t]
    \centering
    \includegraphics[width=0.7\textwidth]{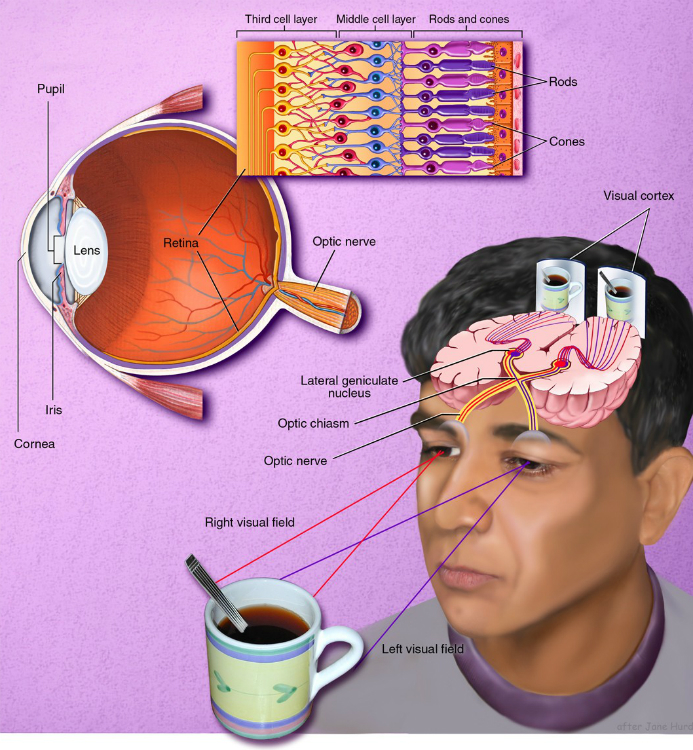}
    \caption{An Illustration of Human Visual System \cite{visual_system_nerve_architecture2}.}
    \label{fig:visual_system}
\end{figure}

As introduced in \cite{visual_system_nerve_architecture2}, vision begins with light passing through the cornea and the lens, which combine to produce a clear image of the visual world on a sheet of photoreceptors called the retina. As in a camera, the image on the retina is reversed: Objects above the center project to the lower part and vice versa. The information from the retina - in the form of electrical signals - is sent via the optic nerve to other parts of the brain, which ultimately process the image and allow us to see. Thus, the visual process begins by comparing the amount of light striking any small region of the retina with the amount of surrounding light. Such a process is also illustrated with Figure~\ref{fig:visual_system}.

Although the visual processing mechanisms are not yet completely understood, recent findings from anatomical and physiological studies in monkeys suggest that visual signals are fed into at least three separate processing systems. One system appears to process information mainly about shape; a second, mainly about color; and a third, movement, location, and spatial organization. Human psychological studies support the findings obtained through animal research. These studies show that the perception of movement, depth, perspective, the relative size of objects, the relative movement of objects, shading, and gradations in texture all depend primarily on contrasts in light intensity rather than on color.

\begin{figure}[t]
    \centering
    \includegraphics[width=0.6\textwidth]{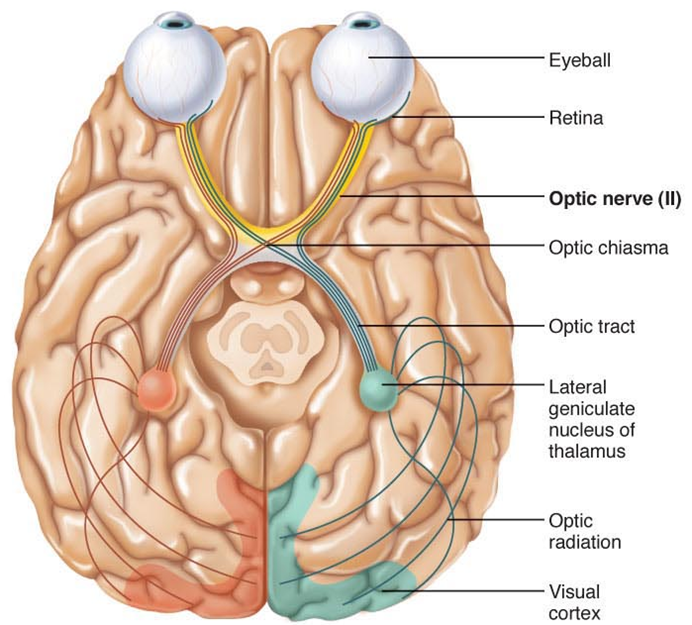}
    \caption{An Illustration of Human Visual System Pathway \cite{visual_system_nerve_architecture}.}
    \label{fig:visual_system_2}
\end{figure}

As shown in Figure~\ref{fig:visual_system_2}, several important biological structures are involved in the visual process, which include the \textit{eye}, \textit{optic nerve}, \textit{optic chiasma}, \textit{optic tract}, \textit{lateral geniculate body}, \textit{optic radiation}, \textit{visual cortes} and \textit{visual association cortex}. These structures form the anterior and posterior pathways in visual process. The anterior visual pathway refers to structures involved in vision before the lateral geniculate nucleus. The posterior visual pathway refers to structures after this point.

\subsection{Auditory System}

\begin{figure}[t]
    \centering
    \includegraphics[width=0.7\textwidth]{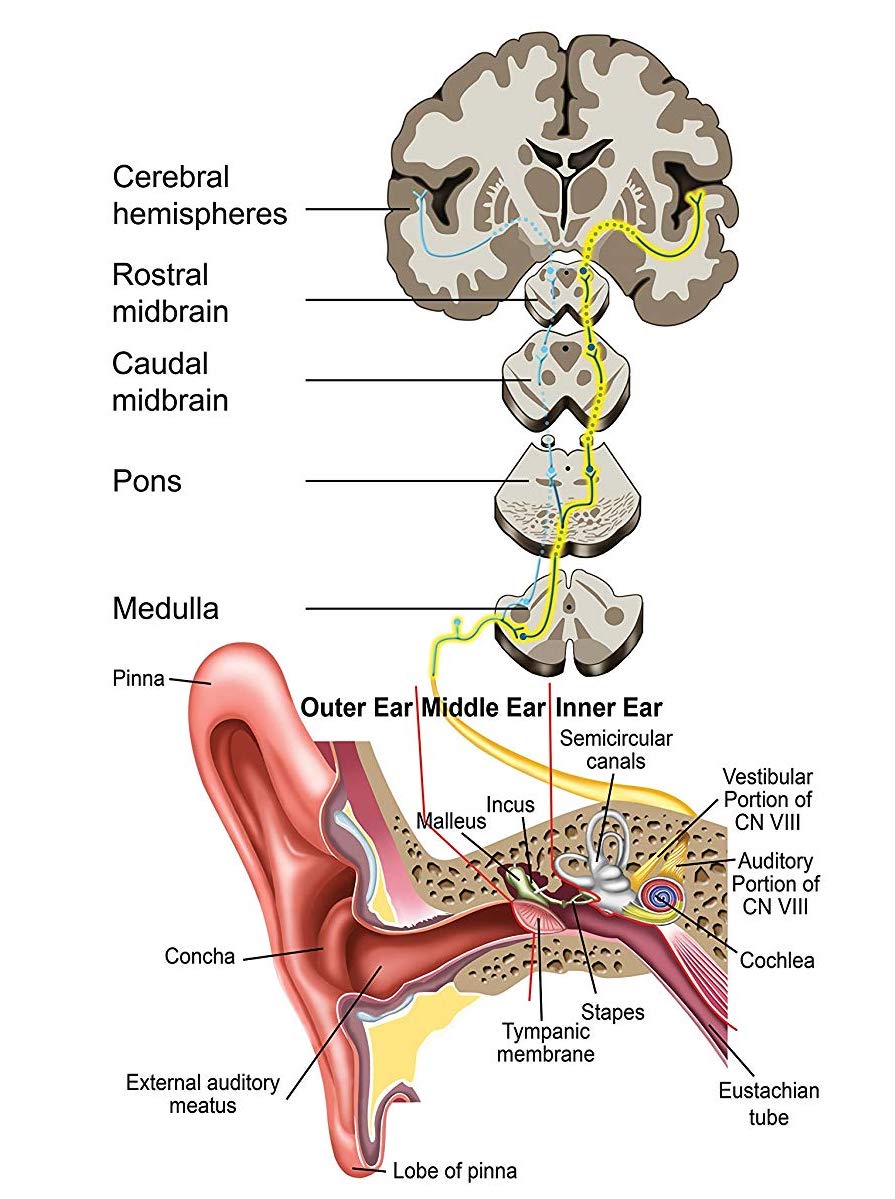}
    \caption{An Illustration of Human Auditory System Pathway \cite{auditory_system}.}
    \label{fig:auditory_pathway}
\end{figure}

The auditory system is the sensory system for the sense of hearing. According to \cite{auditory_system_2}, as illustrated in Figure~\ref{fig:auditory_pathway}, it includes both the peripheral sensory organs (the ears) and the auditory parts of the central sensory system.
\subsubsection{Peripheral Auditory System}

The \textit{peripheral auditory system} includes the \textit{outer ear}, \textit{middle ear} and the \textit{inner ear}.
\begin{itemize}
\item \textbf{Outer Ear}: The pinna are the parts of the outer ear that appear as folds of cartilage. They surround the ear canal and function as sound wave reflectors and attenuators when the waves hit them. The pinna helps the brain identify the direction from where the sounds originated. From the pinna, the sound waves enter a tube-like structure called auditory canal. This canal serves as a sound amplifier. The sound waves travel through the canal and reach the tympanic membrane (eardrum), the canal's end.

\item \textbf{Middle Ear}: As the sound waves hit the eardrum, the sensory information goes into an air-filled cavity through lever-teletype bones called ossicles. The three ossicles include the hammer (malleus), anvil (incus), and stirrup (stapes). These delicate bones convert the sound vibrations made when the sound waves hit the ear drum into sound vibrations of higher pressure. These transformed vibrations (still in wave form) enter the oval window.

\item \textbf{Inner Ear}: Beyond the oval window is the inner ear. This segment of the ear is filled with liquid rather than air, that is why there is a need of conversion of low pressure sound vibrations to higher pressure ones in the middle ear. The main structure in the inner ear is called the cochlea, where the sensory information in wave form is transformed into the neural form. The cochlear duct contains the organ of Corti. This organ is comprised of inner hair cells that turn the vibrations into electric neural signals. Each hair innervates many auditory nerve fibers, and these fibers form the auditory nerve. The auditory nerve (for hearing) combines with the vestibular nerve (for balance), forming cranial nerve VIII or the vestibulocochlear nerve.

\end{itemize}

\subsubsection{Central Auditory System}

Once the sound waves are turned into neural signals, they travel through cranial nerve VIII, reaching different anatomical structures where the neural information is further processed. The cochlear nucleus is the first site of neural processing, followed by the superior olivary complex located in the pons, and then processed in the inferior colliculus at the midbrain. The neural information ends up at the relay center of the brain, called the thalamus. The information is then passed to the primary auditory cortex of the brain, situated in the temporal lobe.

\begin{itemize}

\item \textbf{Primary Auditory Cortex}: The primary auditory cortex receives auditory information from the thalamus. The left posterior superior temporal gyrus is responsible for the perception of sound, and in itthe primary auditory cortex is the region where the attributes of sound (pitch, rhythm, frequency, etc.) are processed.

\end{itemize}

\subsection{Somatosensory System}

The somatosensory system is a part of the sensory nervous system. As introduced in \cite{somatosensory_system}, the somatosensory system is a complex system of sensory neurons and pathways that responds to changes at the surface or inside the body. The axons (as afferent nerve fibers) of sensory neurons connect with, or respond to, various receptor cells. These sensory receptor cells are activated by different stimuli such as heat and nociception, giving a functional name to the responding sensory neuron, such as a thermoreceptor which carries information about temperature changes. Other types include mechanoreceptors, chemoreceptors, and nociceptors which send signals along a sensory nerve to the spinal cord where they may be processed by other sensory neurons and then relayed to the brain for further processing. Sensory receptors are found all over the body including the skin, epithelial tissues, muscles, bones and joints, internal organs, and the cardiovascular system.

\begin{figure}[H]
    \centering
    \includegraphics[width=0.8\textwidth]{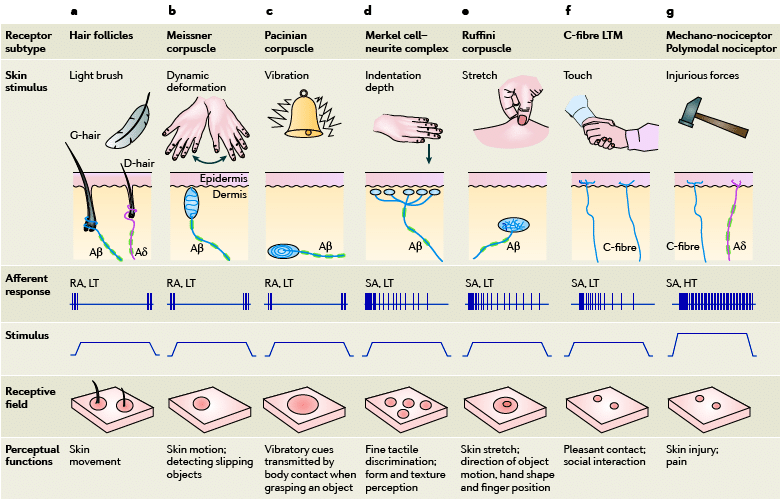}
    \caption{An Illustration of Mechanoreceptors in Skin \cite{mechanoreceptors}.}
    \label{fig:mechanoreceptors}
\end{figure}

According to \cite{mechanoreceptors}, in the skin, there exist several different mechanoreceptors as illustrated in Figure~\ref{fig:mechanoreceptors}, each of which can respond to different stimuli for short or long periods.
\begin{itemize}

\item Hair follicle or root hair plexus is a special group of nerve fiber endings and serves as a very sensitive mechanoreceptor for touch sensation. Each hair plexus forms a network around a hair follicle and is a receptor, which means it sends and receives nerve impulses to and from the brain when the hair moves.

\item Merkel cell nerve endings are found in the basal epidermis and hair follicles; they react to low vibrations (5-15 Hz) and deep static touch such as shapes and edges. Due to having a small receptive field (extremely detailed info), they are used in areas like fingertips the most; they are not covered (shelled) and thus respond to pressures over long periods.

\item Meissner corpuscles react to moderate vibration (10-50 Hz) and light touch. They are located in the dermal papillae; due to their reactivity, they are primarily located in fingertips and lips. They respond in quick action potentials, unlike Merkel nerve endings. They are responsible for the ability to read Braille and feel gentle stimuli.

\item Pacinian corpuscles determine gross touch and distinguish rough and soft substances. They react in quick action potentials, especially to vibrations around 250 Hz (even up to centimeters away). They are the most sensitive to vibrations and have large receptor fields. Pacinian reacts only to sudden stimuli so pressures like clothes that are always compressing their shape are quickly ignored.

\item Ruffini endings react slowly and respond to sustained skin stretch. They are responsible for the feeling of object slippage and play a major role in the kinesthetic sense and control of finger position and movement. Merkel and bulbous cells - slow-response - are myelinated; the rest - fast-response - are not. All of these receptors are activated upon pressures that squish their shape causing an action potential.

\item C-fibre LTM, i.e., the C low-threshold mechanoreceptors (CLTM), which are unmyelinated afferents found in human hairy skin, have a low mechanical threshold $<$5 milliNewtons. They have moderate adaptation and may exhibit fatigue on repetitive stimulation and ``afterdischarges'' for several seconds after a stimulus.

\item Mechano nociceptor, also known as polymodal nociceptors, responds to high intensity stimuli such as mechanical, thermal and to chemical substances, which tend to cause skin injury.

\end{itemize}

\subsection{Taste System}

According to \cite{taste_system}, taste is one of the five traditional senses that belongs to the gustatory system. Taste is the sensation produced when a substance in the mouth reacts chemically with taste receptor cells located on taste buds in the oral cavity, mostly on the tongue. Taste, along with smell (olfaction) and trigeminal nerve stimulation (registering texture, pain, and temperature), determines flavors of food and/or other substances. Humans have taste receptors on taste buds and other areas including the upper surface of the tongue and the epiglottis. The gustatory cortex is responsible for the perception of taste.

\begin{figure}[t]
    \centering
    \includegraphics[width=0.7\textwidth]{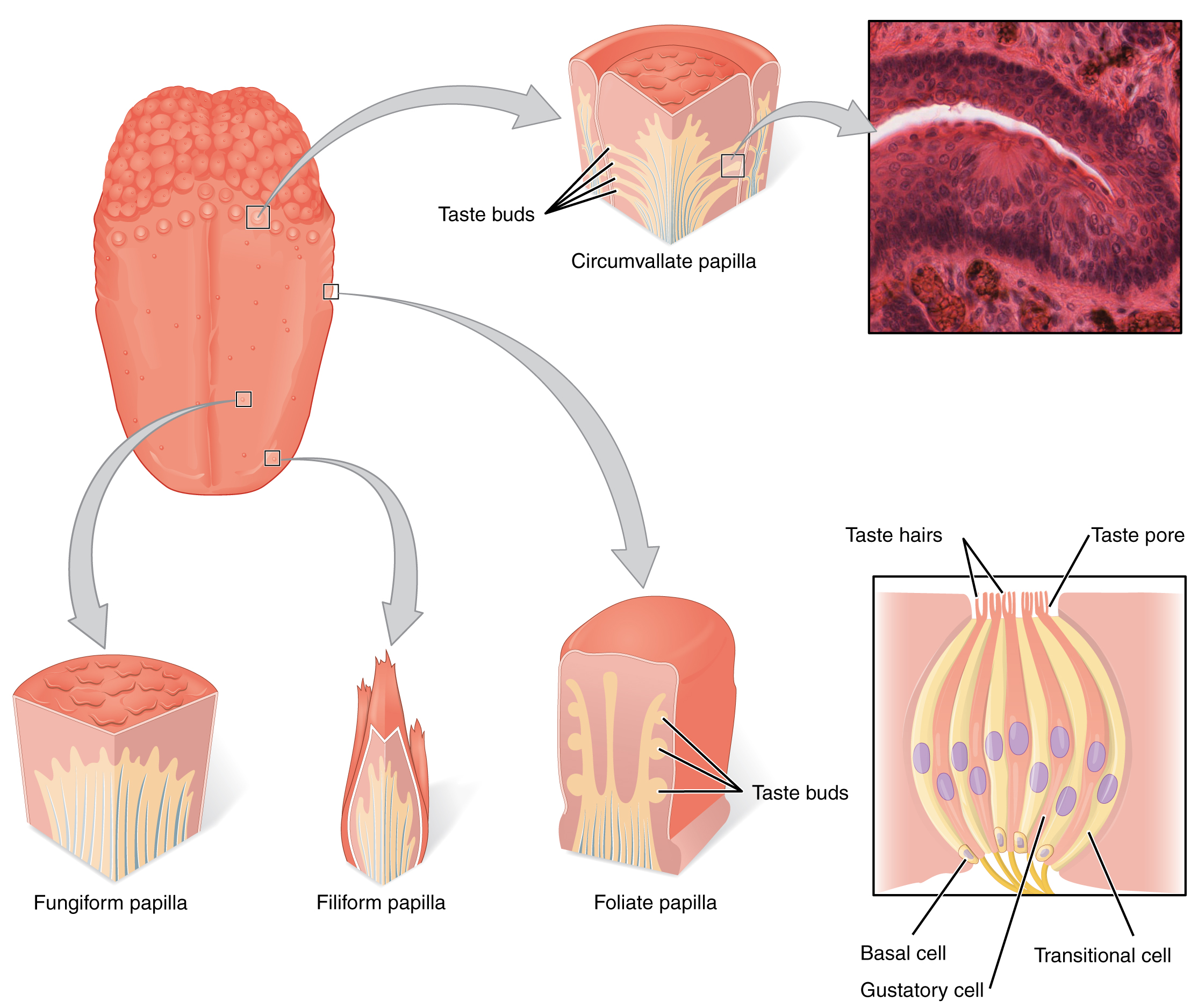}
    \caption{An Illustration of Different Papillae in the Tongue \cite{taste_system}.}
    \label{fig:tongue}
\end{figure}

As illustrated in Figure~\ref{fig:tongue}, the human tongue is covered with thousands of small bumps called papillae, which are visible to the naked eye. Within each papilla are hundreds of taste buds. The exception to this is the filiform papillae that do not contain taste buds. There are between 2000 and 5000 taste buds that are located on the back and front of the tongue. Others are located on the roof, sides and back of the mouth, and in the throat. Each taste bud contains 50 to 100 taste receptor cells, which can differentiate different tastes based on the received stimulus.

In the human body a stimulus refers to a form of energy which elicits a physiological or psychological action or response. Sensory receptors are the structures in the body which change the stimulus from one form of energy to another. This can mean changing the presence of a chemical, sound wave, source of heat, or touch to the skin into an electrical action potential which can be understood by the brain, the body's control center. Sensory receptors are modified ends of sensory neurons; modified to deal with specific types of stimulus, thus there are many different types of sensory receptors in the body. The neuron is the primary component of the nervous system, which transmits messages from sensory receptors all over the body.

Taste is a form of chemoreception which occurs in the specialised taste receptors in the mouth. To date, there are five different types of taste receptors known: salt, sweet, sour, bitter, and umami. Each receptor has a different manner of sensory transduction: that is, of detecting the presence of a certain compound and starting an action potential which alerts the brain. It is a matter of debate whether each taste cell is tuned to one specific tastant or to several; Smith and Margolskee claim that ``gustatory neurons typically respond to more than one kind of stimulus, lthough each neuron responds most strongly to one tastant''. Researchers believe that the brain interprets complex tastes by examining patterns from a large set of neuron responses. This enables the body to make ``keep or spit out'' decisions when there is more than one tastant present. ``No single neuron type alone is capable of discriminating among stimuli or different qualities, because a given cell can respond the same way to disparate stimuli.'' As well, serotonin is thought to act as an intermediary hormone which communicates with taste cells within a taste bud, mediating the signals being sent to the brain. Receptor molecules are found on the top of microvilli of the taste cells.

\begin{itemize}
\item \textbf{Sweetness}: Sweetness is produced by the presence of sugars, some proteins, and a few other substances. It is often connected to aldehydes and ketones, which contain a carbonyl group. Sweetness is detected by a variety of G protein-coupled receptors coupled to a G protein that acts as an intermediary in the communication between taste bud and brain, gustducin. These receptors are T1R2+3 (heterodimer) and T1R3 (homodimer), which account for sweet sensing in humans and other animals.

\item \textbf{Saltness}: Saltiness is a taste produced best by the presence of cations (such as Na+, K+ or Li+) and is directly detected by cation influx into glial like cells via leak channels causing depolarisation of the cell. Other monovalent cations, e.g., ammonium, NH$_4^+$, and divalent cations of the alkali earth metal group of the periodic table, e.g., calcium, Ca$_2^+$, ions, in general, elicit a bitter rather than a salty taste even though they, too, can pass directly through ion channels in the tongue.

\item \textbf{Sourness}: Sourness is acidity, and, like salt, it is a taste sensed using ion channels. Undissociated acid diffuses across the plasma membrane of a presynaptic cell, where it dissociates in accordance with Le Chatelier's principle. The protons that are released then block potassium channels, which depolarise the cell and cause calcium influx. In addition, the taste receptor PKD2L1 has been found to be involved in tasting sour.

\item \textbf{Bitterness}: Research has shown that TAS2Rs (taste receptors, type 2, also known as T2Rs) such as TAS2R38 are responsible for the human ability to taste bitter substances. They are identified not only by their ability to taste certain bitter ligands, but also by the morphology of the receptor itself (surface bound, monomeric).

\item \textbf{Savoriness}: The amino acid glutamic acid is responsible for savoriness, but some nucleotides (inosinic acid and guanylic acid) can act as complements, enhancing the taste. Glutamic acid binds to a variant of the G protein-coupled receptor, producing a savory taste.

\end{itemize}

\subsection{Olfactory System}

As introduced in \cite{olfactory_system}, the olfactory system, or sense of smell, is the part of the sensory system used for smelling (olfaction). Most mammals and reptiles have a main olfactory system and an accessory olfactory system. The main olfactory system detects airborne substances, while the accessory system senses fluid-phase stimuli. The senses of smell and taste (gustatory system) are often referred to together as the chemosensory system, because they both give the brain information about the chemical composition of objects through a process called transduction.

\begin{figure}[t]
    \centering
    \includegraphics[width=0.7\textwidth]{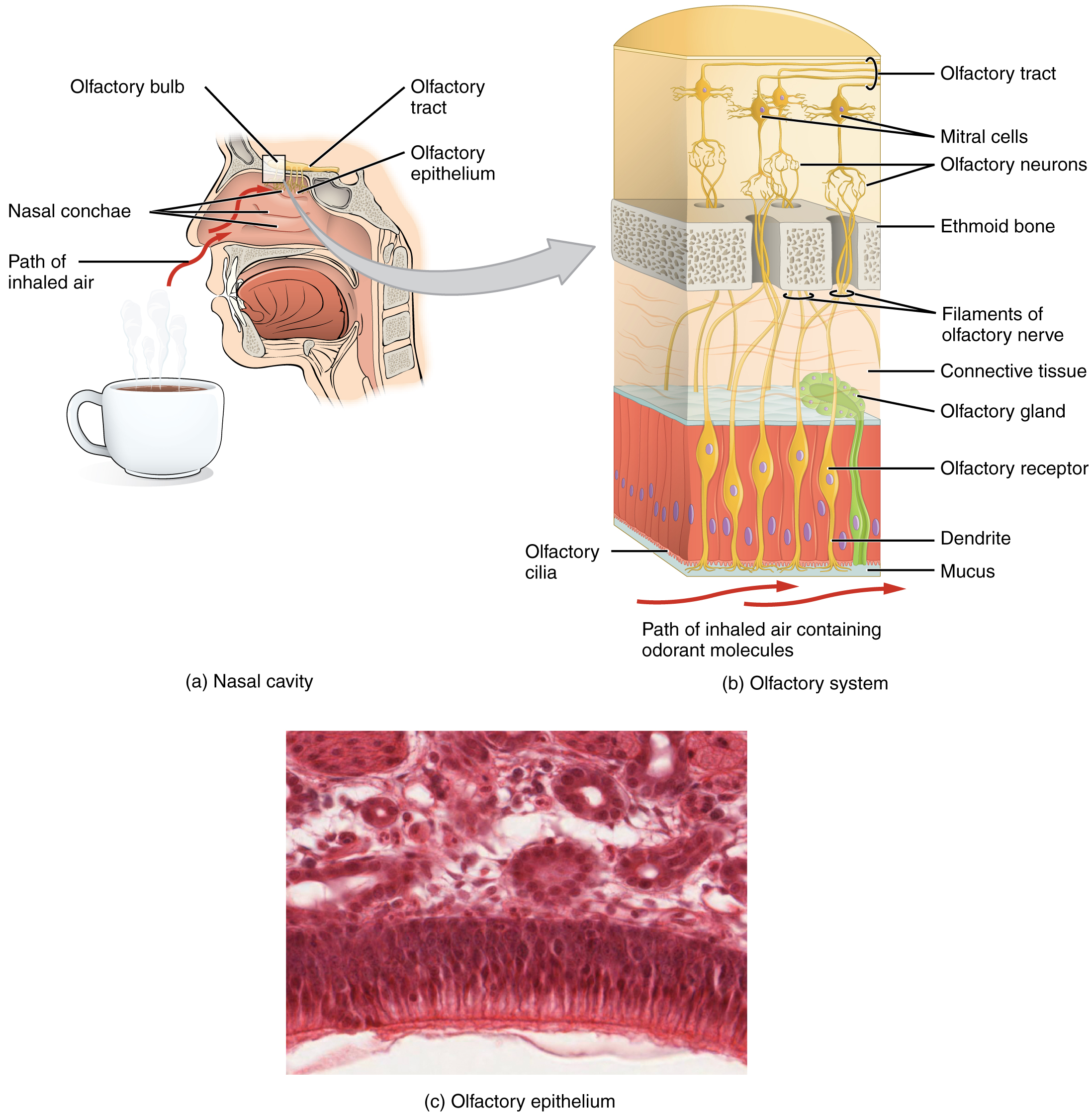}
    \caption{An Illustration of the Peripheral Olfactory System \cite{olfactory_system}.}
    \label{fig:olfactory_system}
\end{figure}

From the structure perspective, the olfactory system includes two main parts: the \textit{peripheral olfactory system} and the \textit{central olfactory system}.

\subsubsection{The Peripheral Olfactory System}

As illustrated in Figure~\ref{fig:olfactory_system}, the peripheral olfactory system consists mainly of the nostrils, ethmoid bone, nasal cavity, and the olfactory epithelium (layers of thin tissue covered in mucus that line the nasal cavity). The primary components of the layers of epithelial tissue are the mucous membranes, olfactory glands, olfactory neurons, and nerve fibers of the olfactory nerves. Odor molecules can enter the peripheral pathway and reach the nasal cavity either through the nostrils when inhaling (olfaction) or through the throat when the tongue pushes air to the back of the nasal cavity while chewing or swallowing (retro-nasal olfaction). Inside the nasal cavity, mucus lining the walls of the cavity dissolves odor molecules. Mucus also covers the olfactory epithelium, which contains mucous membranes that produce and store mucus and olfactory glands that secrete metabolic enzymes found in the mucus.

Olfactory sensory neurons in the epithelium detect odor molecules dissolved in the mucus and transmit information about the odor to the brain in a process called sensory transduction. Olfactory neurons have cilia (tiny hairs) containing Olfactory receptors that bind to odor molecules, causing an electrical response that spreads through the Sensory neuron to the olfactory nerve fibers at the back of the nasal cavity. Olfactory nerves and fibers transmit information about odors from the peripheral olfactory system to the central olfactory system of the brain, which is separated from the epithelium by the cribriform plate of the ethmoid bone. Olfactory nerve fibers, which originate in the epithelium, pass through the cribriform plate, connecting the epithelium to the brain's limbic system at the olfactory bulbs.

\begin{figure}[t]
    \centering
    \includegraphics[width=0.7\textwidth]{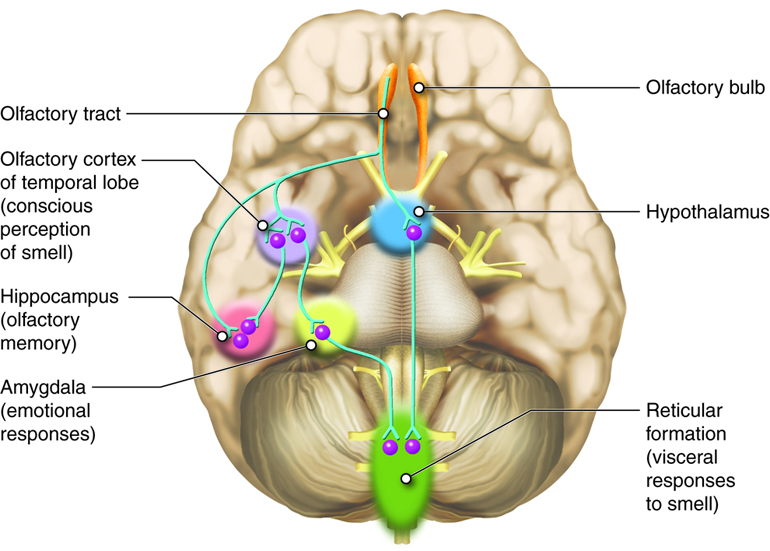}
    \caption{An Illustration of the Central Olfactory System \cite{olfactory_system2}.}
    \label{fig:olfactory_system2}
\end{figure}

\subsubsection{The Central Olfactory System}

As shown in Figure~\ref{fig:olfactory_system2}, the main olfactory bulb transmits pulses to both mitral and tufted cells, which help determine odor concentration based off the time certain neuron clusters fire (called ``timing code''). These cells also note differences between highly similar odors and use that data to aid in later recognition. The cells are different with mitral having low firing-rates and being easily inhibited by neighboring cells, while tufted have high rates of firing and are more difficult to inhibit. The uncus houses the olfactory cortex which includes the piriform cortex (posterior orbitofrontal cortex), amygdala, olfactory tubercle, and parahippocampal gyrus. The olfactory tubercle connects to numerous areas of the amygdala, thalamus, hypothalamus, hippocampus, brain stem, retina, auditory cortex, and olfactory system.

\subsection{Vestibular System}

The vestibular system \cite{vestibular_system}, in most mammals, is the sensory system that provides the leading contribution to the sense of balance and spatial orientation for the purpose of coordinating movement with balance. Together with the cochlea, a part of the auditory system, it constitutes the labyrinth of the inner ear in most mammals. As movements consist of rotations and translations, the vestibular system comprises two components: the semicircular canals which indicate rotational movements; and the otoliths which indicate linear accelerations. The vestibular system sends signals primarily to the neural structures that control eye movements, and to the muscles that keep an animal upright and in general control posture. The projections to the former provide the anatomical basis of the vestibulo-ocular reflex, which is required for clear vision; while the projections to the latter provide the anatomical means required to enable an animal to maintain its desired position in space.

The brain uses information from the vestibular system in the head and from proprioception throughout the body to enable the animal to understand its body's dynamics and kinematics (including its position and acceleration) from moment to moment. How these two perceptive sources are integrated to provide the underlying structure of the sensorium is unknown. Next, we will briefly introduce the \textit{semicircular canal system} and the \textit{otolithic organs}, which compose the vestibular system.

\subsubsection{Semicircular Canal System}

\begin{figure}[t]
    \centering
    \includegraphics[width=0.8\textwidth]{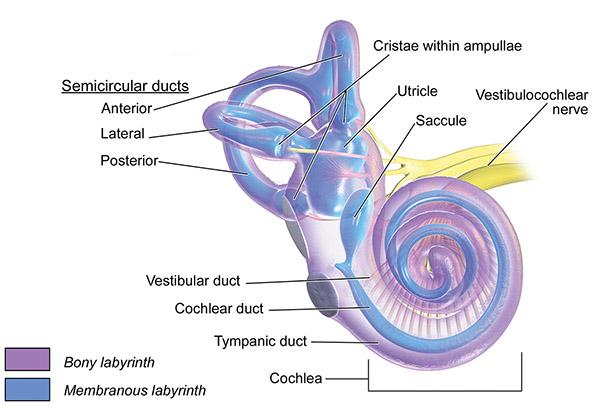}
    \caption{An Illustration of the Semicircular Canal System \cite{semicircular_canal_system}.}
    \label{fig:semicircular_canal_system}
\end{figure}

Since the world is three-dimensional, the vestibular system contains three semicircular canals in each labyrinth, as illustrated in Figure~\ref{fig:semicircular_canal_system}. They are approximately orthogonal (at right angles) to each other, and are the horizontal (or lateral), the anterior semicircular canal (or superior), and the posterior (or inferior) semicircular canal. Anterior and posterior canals may collectively be called vertical semicircular canals. Movement of fluid within the horizontal semicircular canal corresponds to rotation of the head around a vertical axis (i.e. the neck), as when doing a pirouette. The anterior and posterior semicircular canals detect rotations of the head in the sagittal plane (as when nodding), and in the frontal plane, as when cartwheeling. Both anterior and posterior canals are orientated at approximately 45$^\circ$ between frontal and sagittal planes.

The canals are arranged in such a way that each canal on the left side has an almost parallel counterpart on the right side. Each of these three pairs works in a push-pull fashion: when one canal is stimulated, its corresponding partner on the other side is inhibited, and vice versa. This push-pull system makes it possible to sense all directions of rotation: while the right horizontal canal gets stimulated during head rotations to the right, the left horizontal canal gets stimulated (and thus predominantly signals) by head rotations to the left.

\subsubsection{Otolithic Organs}

While the semicircular canals respond to rotations, the otolithic organs sense linear accelerations. Humans have two otolithic organs on each side, one called the utricle, the other called the saccule. The utricle contains a patch of hair cells and supporting cells called a macula. Similarly, the saccule contains a patch of hair cells and a macula. Each hair cell of a macula has 40-70 stereocilia and one true cilium called a kinocilium. The tips of these cilia are embedded in an otolithic membrane. This membrane is weighted down with protein-calcium carbonate granules called otoliths. These otoliths add to the weight and inertia of the membrane and enhance the sense of gravity and motion. With the head erect, the otolithic membrane bears directly down on the hair cells and stimulation is minimal. When the head is tilted, however, the otolithic membrane sags and bends the stereocilia, stimulating the hair cells. Any orientation of the head causes a combination of stimulation to the utricles and saccules of the two ears. The brain interprets head orientation by comparing these inputs to each other and to other input from the eyes and stretch receptors in the neck, thereby detecting whether the head is tilted or the entire body is tipping. Essentially, these otolithic organs sense how quickly you are accelerating forward or backward, left or right, or up or down. Most of the utricular signals elicit eye movements, while the majority of the saccular signals projects to muscles that control our posture.

While the interpretation of the rotation signals from the semicircular canals is straightforward, the interpretation of otolith signals is more difficult: since gravity is equivalent to a constant linear acceleration, one somehow has to distinguish otolith signals that are caused by linear movements from those caused by gravity. Humans can do that quite well, but the neural mechanisms underlying this separation are not yet fully understood. Humans can sense head tilting and linear acceleration even in dark environments because of the orientation of two groups of hair cell bundles on either side of the striola. Hair cells on opposite sides move with mirror symmetry, so when one side is moved, the other is inhibited. The opposing effects caused by a tilt of the head cause differential sensory inputs from the hair cell bundles allow humans to tell which way the head is tilting, Sensory information is then sent to the brain, which can respond with appropriate corrective actions to the nervous and muscular systems to ensure that balance and awareness are maintained.

\section{Summary}

In this article, we have introduced the brain biological structure and functions, and we also introduce its several important sensory systems, which receive information from the environment for the brain. In the follow-up two articles, i.e., \cite{zhang2019basic} and \cite{zhang2019cognitive}, we will further introduce the low-level composition basis structures (e.g., neuron, synapse and action potential) and the high-level cognitive functions (e.g., consciousness, attention, learning and memory) of the brain, respectively.

%
%
%

%
%
%

\newpage

\vskip 0.2in
\bibliographystyle{plain}
\bibliography{reference}

\begin{thebibliography}{10}

\bibitem{auditory_system}
Auditory system.
\newblock \url{https://en.wikipedia.org/wiki/Auditory_system}.
\newblock [Online; accessed 30-April-2019].

\bibitem{auditory_system_2}
The auditory system.
\newblock \url{https://explorable.com/auditory-system}.
\newblock [Online; accessed 30-April-2019].

\bibitem{brain_structure}
Brain.
\newblock \url{https://en.wikipedia.org/wiki/Brain}.
\newblock [Online; accessed 22-April-2019].

\bibitem{brain_stem}
Brainstem.
\newblock \url{https://en.wikipedia.org/wiki/Brainstem}.
\newblock [Online; accessed 23-April-2019].

\bibitem{brodmann_area}
Brodmann area.
\newblock \url{https://en.wikipedia.org/wiki/Brodmann_area}.
\newblock [Online; accessed 30-April-2019].

\bibitem{brodmann_area_plot}
Brodmann areas.
\newblock \url{https://www.kenhub.com/en/library/anatomy/brodmann-areas}.
\newblock [Online; accessed 30-April-2019].

\bibitem{brain_comparison}
Concept 34.5: Neurons are organized into nervous systems.
\newblock
  \url{http://www.macmillanhighered.com/BrainHoney/Resource/6716/digital_first_content/trunk/test/hillis2e/hillis2e_ch34_6.html}.
\newblock [Online; accessed 22-April-2019].

\bibitem{visual_system_nerve_architecture}
Cranial nerve \#2: Optic nerve.
\newblock
  \url{https://12cranialnerves.wordpress.com/cranial-nerve-2-optic-nerve/}.
\newblock [Online; accessed 30-April-2019].

\bibitem{brain_diencephalon}
The diencephalon.
\newblock \url{https://antranik.org/the-diencephalon/}.
\newblock [Online; accessed 22-April-2019].

\bibitem{nervous_system}
Evolution of nervous systems.
\newblock
  \url{https://content.openclass.com/eps/pearson-reader/api/item/ab914c98-1923-486b-bdb4-b9187be18b9e/1/file/silverthornHP7-071415-MJ-BO/OPS/s9ml/chapter09/filep700049593400000000000000000353e.xhtml}.
\newblock [Online; accessed 22-April-2019].

\bibitem{brain_diencephalon_description}
Know your brain: Diencephalon.
\newblock
  \url{https://neuroscientificallychallenged.com/blog/know-your-brain-diencephalon}.
\newblock [Online; accessed 23-April-2019].

\bibitem{semicircular_canal_system}
Listen to your ears.
\newblock
  \url{https://archive.divernet.com/medical-health/p315302-listen-to-your-ears.html}.
\newblock [Online; accessed 30-April-2019].

\bibitem{olfactory_system}
Olfactory system.
\newblock \url{https://en.wikipedia.org/wiki/Olfactory_system}.
\newblock [Online; accessed 30-April-2019].

\bibitem{pituitary_gland}
Pituitary gland.
\newblock \url{https://en.wikipedia.org/wiki/Pituitary_gland}.
\newblock [Online; accessed 23-April-2019].

\bibitem{somatosensory_system}
Somatosensory system.
\newblock \url{https://en.wikipedia.org/wiki/Somatosensory_system}.
\newblock [Online; accessed 30-April-2019].

\bibitem{olfactory_system2}
Special senses: Smell (olfaction).
\newblock
  \url{https://courses.lumenlearning.com/austincc-ap1/chapter/special-senses-smell-olfaction/}.
\newblock [Online; accessed 30-April-2019].

\bibitem{taste_system}
Taste.
\newblock \url{https://en.wikipedia.org/wiki/Taste}.
\newblock [Online; accessed 30-April-2019].

\bibitem{vestibular_system}
Vestibular system.
\newblock \url{https://en.wikipedia.org/wiki/Vestibular_system}.
\newblock [Online; accessed 30-April-2019].

\bibitem{visual_system_nerve_architecture2}
Vision: Processing information.
\newblock
  \url{https://www.brainfacts.org/Thinking-Sensing-and-Behaving/Vision/2012/Vision-Processing-Information}.
\newblock [Online; accessed 30-April-2019].

\bibitem{visual_system}
Visual system.
\newblock \url{https://en.wikipedia.org/wiki/Visual_system}.
\newblock [Online; accessed 30-April-2019].

\bibitem{phdthesis}
Amir Ahangi.
\newblock {\em Using Combining Classifiers in Brain-Computer Interfacing}.
\newblock PhD thesis, 02 2012.

\bibitem{pituitary_gland_plot}
Regina Bailey.
\newblock Pituitary gland.
\newblock \url{https://www.thoughtco.com/pituitary-gland-anatomy-373226}.
\newblock [Online; accessed 23-April-2019].

\bibitem{brain_cerebellum}
M.D. J.~Keith~Fisher.
\newblock Everything you need to know about the cerebellum.
\newblock \url{https://www.medicalnewstoday.com/articles/313265.php}.
\newblock [Online; accessed 23-April-2019].

\bibitem{brain_amatomy}
Kayt Sukel.
\newblock Neuroanatomy-a primer(1).
\newblock \url{http://www.dana.org/News/Details.aspx?id=43515}.
\newblock [Online; accessed 22-April-2019].

\bibitem{mechanoreceptors}
Thomas Suslak.
\newblock {\em There and back again: a stretch receptor's tale}.
\newblock PhD thesis, 06 2015.

\bibitem{zhang2019basic}
Jiawei Zhang.
\newblock Basic neural units of the brain: Neurons, synapses and action
  potential, 2019.

\bibitem{zhang2019cognitive}
Jiawei Zhang.
\newblock Cognitive functions of the brain: Perception, attention and memory,
  2019.

\end{thebibliography}

\end{document}